\documentclass[journal]{IEEEtran}
\usepackage{amsmath,amsfonts}
\usepackage{algorithmic}
\usepackage{array}
\usepackage[caption=false,font=normalsize,labelfont=sf,textfont=sf]{subfig}
\usepackage{textcomp}
\usepackage{stfloats}
\usepackage{url}
\usepackage{verbatim}
\usepackage{graphicx}
\usepackage{bm}
\usepackage{multirow}
\usepackage{makecell}
\usepackage{booktabs}
\usepackage{color}
\usepackage{adjustbox}
\usepackage{cite}
\usepackage{xcolor}
\usepackage{colortbl}
\usepackage{algorithm,algorithmic}
\usepackage{threeparttable}
\usepackage{balance}
\begin{document}
\bstctlcite{settingbib}

\title{WiFo-CF: Wireless Foundation Model for  \\ CSI Feedback}
\author{\IEEEauthorblockN{
		Xuanyu Liu,~\IEEEmembership{Graduate Student Member,~IEEE,}~Shijian Gao,~\IEEEmembership{Member,~IEEE,}~Boxun Liu,~\IEEEmembership{Graduate Student Member,~IEEE,}~Xiang Cheng,~\IEEEmembership{Fellow,~IEEE,}~and Liuqing Yang,~\IEEEmembership{Fellow,~IEEE}
        }
\thanks{
Xuanyu Liu, Boxun Liu and Xiang Cheng are with the State Key Laboratory of Photonics and Communications, School of Electronics, Peking University, Beijing 100871, China (e-mail: xyliu25@stu.pku.edu.cn; boxunliu@stu.pku.edu.cn; xiangcheng@pku.edu.cn).
			
Shijian Gao is with the Internet of Things Thrust, The Hong Kong University of Science and Technology (Guangzhou), Guangzhou 511400, China (e-mail: shijiangao@hkust-gz.edu.cn).
		
Liuqing Yang is with the Internet of Things Thrust and Intelligent Transportation Thrust, The Hong Kong University of Science and Technology (Guangzhou), Guangzhou 511400, China, and also with the Department of Electronic and Computer Engineering and the Department of Civil and Environmental Engineering, The Hong Kong University of Science and Technology, Hong Kong, SAR, China (e-mail: lqyang@ust.hk).
}} 

\maketitle

\begin{abstract}

Deep learning-based channel state information (CSI) feedback schemes demonstrate strong compression capabilities but are typically constrained to fixed system configurations, limiting their generalization and flexibility. To address this challenge, WiFo-CF—a novel wireless foundation model tailored for CSI feedback—is proposed, uniquely accommodating heterogeneous configurations such as varying channel dimensions, feedback rates, and data distributions within a unified framework through its key innovations: (1) a multi-user, multi-rate self-supervised pre-training strategy; and (2) a Mixture of Shared and Routed Expert (S-R MoE) architecture. Supporting the large-scale pre-training of WiFo-CF is the first heterogeneous channel feedback dataset, whose diverse patterns enable the model to achieve superior performance on both in-distribution and out-of-distribution data across simulated and real-world scenarios. Furthermore, the learned representations effectively facilitate adaptation to downstream tasks such as CSI-based indoor localization, validating WiFo-CF's scalability and deployment potential.


\end{abstract}


\section{Introduction}

Massive multiple-input multiple-output (MIMO) is a fundamental enabling technology in 5G \cite{larsson2014massive, lu2014overview, rusek2012scaling, cheng2023intelligent, liu2025beam}, which significantly improves spectral efficiency by deploying a large number of antennas at the base station (BS) and employing high-resolution beamforming techniques.
Accurate channel state information (CSI) is essential for the effective deployment of m-MIMO systems, as it underpins critical functions such as transceiver design \cite{zhang2024integrated}, adaptive modulation \cite{chung2001degrees}, and resource allocation \cite{sadr2009radio}.
In time division duplexing (TDD) systems, the BS can directly estimate the downlink channel by leveraging channel reciprocity and uplink pilot transmissions. However, in frequency division duplexing (FDD) systems, due to the separation of uplink and downlink frequency bands, such reciprocity does not hold. Consequently, the downlink CSI must be estimated at the user equipment (UE) and fed back to the BS via the uplink, which incurs substantial communication overhead. This feedback burden not only leads to inefficient utilization of uplink resources but also scales unfavorably with the increasing number of antennas and subcarriers.

\textcolor{black}{
To effectively reduce the feedback overhead, existing works can be categorized into three approaches: codebook-based schemes, compressive sensing (CS)-based schemes, and deep learning (DL)-based schemes. Codebook-based schemes tend to have low accuracy and a feedback cost that grows linearly with the channel matrix size \cite{huang2009limited, kim2012differential}. Moreover, related studies have leveraged the sparsity of the channel matrix by developing CS-based methods, including LASSO \cite{daubechies2004iterative}, BM3D-AMP \cite{metzler2016denoising}, and TVAL3 \cite{li2009user} for efficient channel compression, and DSDS \cite{gao2020estimating} for path extraction. Nevertheless, the sparsity is often unreliable in practice, and iterative decoders introduce nontrivial base-station latency. In recent years, DL-based approaches have shown greater promise. The pioneering work \cite{wen2018deep} introduced CsiNet, an autoencoder framework capable of compressing and reconstructing CSI, and \cite{guo2020convolutional} proposed CsiNet Plus to address quantization effects encountered in practical feedback systems. Moreover, subsequent investigations such as \cite{wang2018deep, lu2020multi, cai2019attention, ji2021clnet, mashhadi2020distributed} have refined network architectures and harnessed channel correlations to further elevate feedback performance. Despite these advances, limited training data and bounded model capacity still constrain robustness and generalization in highly dynamic, complex scenarios.
}

\textcolor{black}{
With increasingly complex and dynamic channel environments, existing lightweight models suffer from limited representational capacity, resulting in marked performance degradation \cite{zhang2024zone}. When the underlying channel distribution shifts, these models must be retrained with newly collected CSI, yet gathering full CSI in MIMO-OFDM imposes substantial communication and hardware costs. Furthermore, practical wireless systems span diverse frequency bands (sub-6 GHz, mmWave, THz), deployment scenarios (UMa, UMi, RMa, indoor), antenna geometries (linear, planar, elliptical), and even equipment from different vendors, thereby generating CSI with highly heterogeneous dimensions and statistics. Existing deep-learning solutions cannot accommodate this diversity within a single model; instead, they require configuration-specific training and deployment, which not only raises overhead but also prevents the exploitation of cross-configuration correlations that could further enhance performance.
}

The emergence of foundation models \cite{liu2024llm4cp,liu2025llm4wm,cheng2025foundation, han2025llm4sp} provides an effective pathway to address these challenges. 
LWM \cite{alikhani2024large} leverages unsupervised pre-training to learn robust and universal channel feature representations, thereby empowering a wide range of downstream tasks. WiFo \cite{liu2024wifo}, on the other hand, is the first to propose a pre-training strategy specifically tailored for channel prediction tasks, demonstrating strong zero-shot generalization capabilities.  
\textcolor{black}{
Similarly, \cite{emery2025foundation} builds a precoding foundation model via unsupervised pre-training and achieves excellent zero-shot and few-shot performance. However, none of these approaches can be directly applied to CSI-feedback tasks.
Focusing on CSI feedback, \cite{guo2025prompt} introduces LAM, a framework designed to improve generalization by jointly training on datasets with diverse user distributions. However, LAM remains limited in handling heterogeneous system configurations. Specifically, it assumes a fixed input shape, making it inflexible to variations in antenna count, user number, or subcarrier allocation. This structural rigidity not only hinders adaptability but also restricts pre-training to datasets with uniform configurations, thereby reducing data diversity and increasing the risk of overfitting.
}
 
 

Considering that existing works still struggle to handle the aforementioned heterogeneous CSI feedback, our objective is to extend the paradigm of wireless foundation models to the CSI feedback task. By conducting unsupervised pre‐training on a large‐scale heterogeneous channel dataset, we enable the model to acquire a robust and universal capability for channel compression and reconstruction. As a result, it can effectively address heterogeneous channel feedback, achieve strong generalization, and empower downstream channel-related tasks.
Unlike existing small models tailored for specific scenarios, we propose a wireless foundation model for CSI feedback (WiFo-CF), enabling unified handling of heterogeneous configurations. Specifically, on one hand, we design a flexible network architecture to process heterogeneous channel data. To handle heterogeneity in channel dimensions, we draw inspiration from WiFo’s 3D CSI patching operation and extend it to the multi-user case, thus accommodating a dynamic number of users. To address heterogeneity in channel distributions, we introduce an Mixture of Shared and Routed Expert (S-R MoE) architecture that fully extracts both correlations and unique characteristics across heterogeneous datasets, leveraging the advantages of joint training on diverse data. On the other hand, we devise a dedicated self-supervised pre-training strategy so that the model can learn universal channel representations targeted at channel compression and reconstruction. Recognizing that CSI feedback is inherently a self-supervised task, we design a multi-user, multi-rate CSI feedback pre-training task, allowing the model to directly learn generalizable CSI feedback knowledge and significantly improving its zero-shot generalization capability.
The core contributions of this paper are summarized below:

\begin{itemize}

\item We introduce the first wireless foundation model for heterogeneous CSI feedback, capable of flexibly managing CSI feedback tasks under a variety of configurations, including feedback rate, CSI dimensions, user distributions, and cell scenarios.

\item We propose a multi-rate, multi-user joint pre-training strategy and design an S-R MoE architecture to facilitate universal channel representations. This architecture is engineered to extract both common correlations and unique characteristics across diverse datasets.

\item We create a large-scale CSI feedback dataset using statistical models, ray-tracing, and real-world measurements. Our model shows strong performance both in-distribution and out-of-distribution, with fine-tuning experiments confirming the effectiveness of the universal channel representations from WiFo-CF.

\end{itemize}

\textit{Notation:} $(\cdot)^{\rm H}$, $\lvert \cdot \rvert$ and $\Vert\cdot\Vert$ denote the conjugate transpose, determinant and $l_2$ norm, respectively. $\bm{a}[i]$ is the $i\mbox{-}$th element of a vector $\bm{a}$ and $\bm{M}[i,j]$ denotes the element of matrix or tensor $\bm{M}$ at the $i\mbox{-}$th row and the $j\mbox{-}$th column.
$\mathbb{E}\{\cdot\}$ denotes the statistical expectation of the enclosed variable or expression.

\section{System Model and Problem Formulation}

\subsection{System Description}

We consider a frequency division duplexing (FDD) multi-user multiple-input single-output orthogonal frequency division multiplexing (MU-MIMO-OFDM) system, where the base station (BS) is equipped with \( N_t \) antennas and simultaneously serves \( K \) single-antenna user equipments (UEs). The OFDM system consists of \( N_c \) subcarriers, with \( \Delta f \) denoting the subcarrier spacing.
For each subcarrier index \( n \in \{1, \dots, N_c\} \), the BS applies linear precoding to transmit user-specific data symbols \( s_{k,n} \). Thus, the transmit signal on \( n \)-th subcarrier can be written as:
\begin{equation}
\label{V_ofdm}
\mathbf{x}_n = \sum_{k=1}^K \mathbf{v}_{k,n} s_{k,n} = \mathbf{V}_n \mathbf{s}_n,
\end{equation}
where \( \mathbf{v}_{k,n} \in \mathbb{C}^{N_t} \) denotes the precoding vector for the \( k \)-th user on \( n \)-th subcarrier , which is also the \( k \)-th column of the precoding matrix \( \mathbf{V}_n \in \mathbb{C}^{N_t \times K} \). The precoding matrix satisfies a total power constraint per subcarrier as \( \text{Tr}(\mathbf{V}_n \mathbf{V}_n^H) \leq P \), and the transmitted symbol vector \( \mathbf{s}_n \in \mathbb{C}^{K} \) is assumed to be normalized such that \( \mathbb{E}[\mathbf{s}_n \mathbf{s}_n^H] = \mathbf{I} \).

The classical cluster-based multipath channel model is utilized to describe the downlink and uplink CSI between the BS and UE at frequency $f$:
\begin{equation}
\resizebox{0.9\hsize}{!}{$
\begin{aligned}\label{CSI}  
&\bm{h}(f)=\sum_{m=1}^M \sum_{p=1}^{P_m}\beta_{m,p}e^{-j2\pi f_0\tau_{m,p}}e^{j\Phi_{m,p}}\bm{a}(\theta_{m,p}, \phi_{m,p}).
\end{aligned}$}
\end{equation}
In this context, $M$ and $P_m$ are the number of clusters and paths in each cluster, respectively.
$\beta_{m,p}$, $\tau_{m,p}$, and $\Phi_{m,p}$ represent the complex path gain, delay, and random phase, respectively.
$\bm{a}(\theta_{m,p}, \phi_{m,p})$ represents the steering vector of the corresponding path, where $\theta_{m,p}$ and $\phi_{m,p}$ denote the azimuth and elevation angles, respectively.

Let \( \mathbf{h}_{k, n} \in \mathbb{C}^{N_t} \) denote the wireless channel between the BS and the $k$-th UE on $n$-th subcarrier, which satisfies \( \mathbf{h}_{k, n} = h(f_1 + (n - 1)\times \Delta f), \) where $f_1$ denoted the frequency of the first subcarrier. 
Thus, for each subcarrier index \( f \), the received signal at user \( k \) can be given by:
\begin{equation}
y_{k,n} = \mathbf{h}_{k,n}^H \mathbf{v}_{k,n} s_{k,n} + \sum_{i \neq k} \mathbf{h}_{k,n}^H \mathbf{v}_{i,n} s_{i,n} + n_{k,n},
\end{equation}
where \( n_{k,n} \sim \mathcal{CN}(0, \sigma^2) \) denotes additive white Gaussian noise.

Based on the above received signal model, the achievable rate for user \( k \) is given by summing the per-subcarrier rate over all subcarriers:
\begin{equation}
\label{Rk_OFDM}
R_k = \sum_{n=1}^{N_c} \log_2 \left(1 + \frac{|\mathbf{h}_{k,n}^H \mathbf{v}_{k,n}|^2}{\sum_{i \neq k} |\mathbf{h}_{k,n}^H \mathbf{v}_{i,n}|^2 + \sigma^2} \right).
\end{equation}

To eliminate inter-user interference, the Zero-forcing (ZF) precoding \cite{joham2005linear} is applied as
\begin{equation}
\mathbf{V}_n = \gamma \mathbf{H}_n^H(\mathbf{H}_n \mathbf{H}_n^H)^{-1},  
\end{equation}
\textcolor{black}{
where $\gamma$ is a scalar chosen to satisfy the transmit-power constraint, $\mathbf{H}_n = [\mathbf{h}_{n,1}^{H}, ..., \mathbf{h}_{n, N_c}^{H}] \in \mathbb{C}^{K \times N_t}$. 
Advanced designs, such as WMMSE \cite{shi2011iteratively} or learning-based precoders, are omitted, as this work focuses exclusively on CSI feedback rather than precoder optimization.
}

\begin{figure*}[htbp]
    \centering
    \vspace{-1em}
    \includegraphics[width=0.9\linewidth]{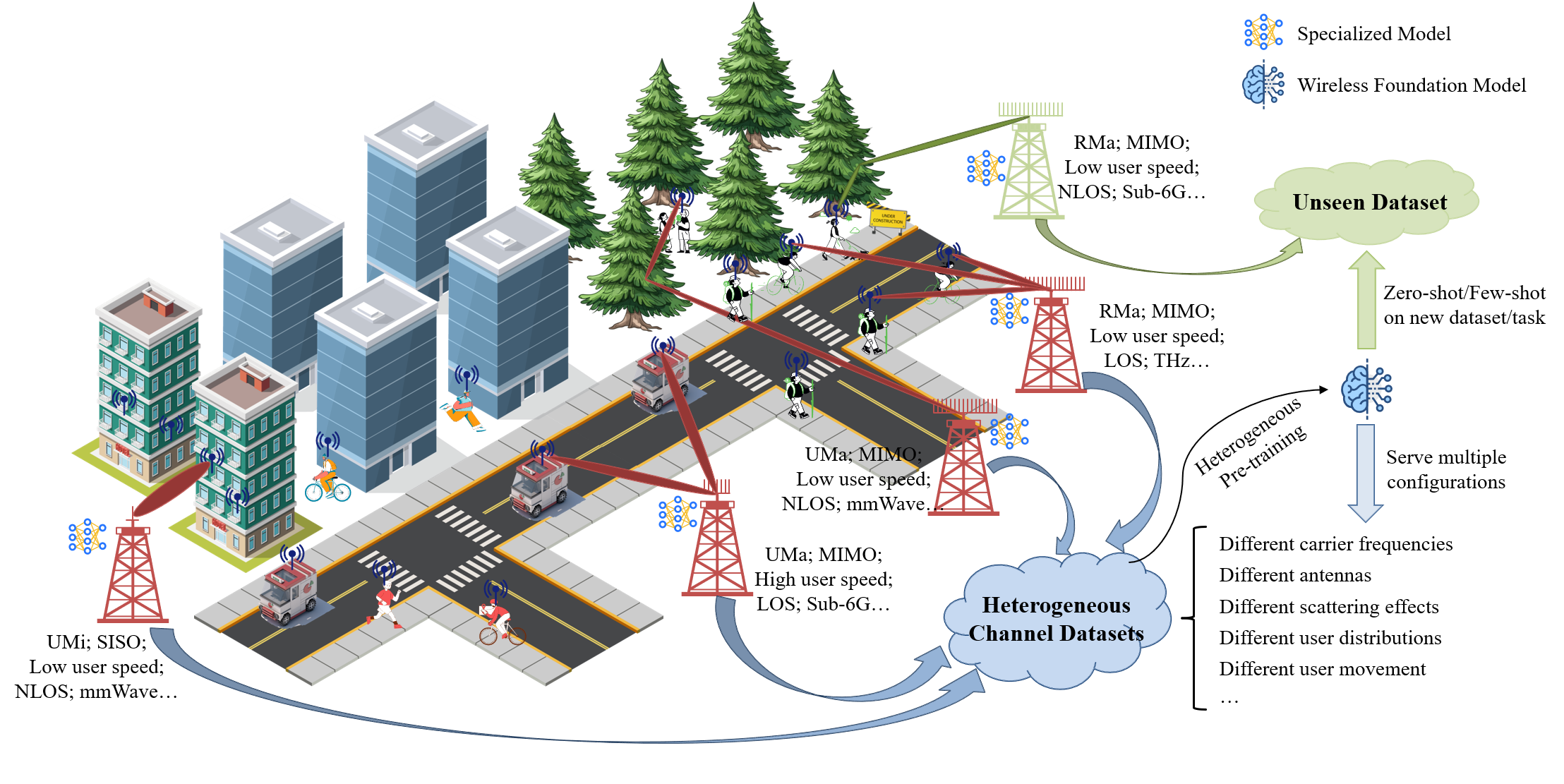}
    \caption{An illustration that highlights the differences in workflows between specialized models and the wireless foundation model.}
    \label{workflow}
    \vspace{-1em}
    \end{figure*}

\subsection{Heterogeneous CSI feedback}

In a multi-user FDD system, each user first estimates its downlink channel matrix, denoted by \(\mathbf{H}_{k} \in \mathbb{C}^{N_c \times N_t}\), via downlink pilot signals. To reduce the feedback overhead, each user would employ an encoder network to compress the high-dimensional channel matrix into a low-dimensional representation. This encoding process can be expressed as:
\begin{equation}
\mathbf{v}_{k} = \mathcal{E}_{k}\bigl(\mathbf{H}_{k};\,\boldsymbol{\theta}_{\mathcal{E},k}\bigr),
\end{equation}
where \(\mathcal{E}_{k}(\cdot)\) denotes the encoder deployed at user \(k\), and \(\boldsymbol{\theta}_{\mathcal{E},k}\) represents its network parameters. The output \(\mathbf{v}_{k} \in \mathbb{R}^{M}\) is a latent feature vector of dimension \(D\). We define the compression ratio \(\mathrm{CR}=\frac{D}{2N_{t}N_{c}}\). This ratio quantifies the reduction in feedback overhead achieved by the encoder.

To enable uplink transmission of the compressed features, \(\mathbf{v}_{k}\) is quantized using a uniform \(b\)-bit quantizer. The quantized codeword is given by
\begin{equation}
\mathbf{s}_{k} = \mathcal{Q}\bigl(\mathbf{v}_{k}) \in \{+1,\,-1\}^{bD},
\end{equation}
where $\mathcal{Q}(\cdot)$ denotes the quantization function, which adopts a $b$-bit non-uniform $\mu$-law quantizer. Consequently, the total number of feedback bits per user is \( N_{\text{bit}} = bD = 2b\,\mathrm{CR}\,N_{t}N_{c} \).

Upon receiving the quantized codewords \(\{\mathbf{s}_{k}\}_{k=1}^{K}\) from all \(K\) users, the BS performs dequantization and decoding to reconstruct the complete multi-user CSI:
\begin{equation}
\bigl[\mathbf{H}_{1},\,\mathbf{H}_{2},\,\ldots,\,\mathbf{H}_{K}\bigr]
= \mathcal{R}\bigl(\bigl\{      \mathcal{D}(\mathbf{s}_{k})\bigr\}_{k=1}^{K};\,\boldsymbol{\theta}_{\mathcal{R}}\bigr),
\end{equation}
where \(\mathcal{R}(\cdot)\) denotes the decoder function parameterized by \(\boldsymbol{\theta}_{\mathcal{R}}\), \(\mathcal{D}(\cdot)\) denotes the dequantization function.

However, in practical deployments, channel statistics and propagation environments vary substantially across different frequency bands, user mobility levels, antenna configurations, and spatial distributions. Such variations give rise to heterogeneous CSI distributions, which challenge traditional feedback models trained under a single fixed data distribution.

To address this challenge, we consider a \emph{heterogeneous CSI feedback} problem, where the objective is to train a unified network that can achieve high-fidelity CSI reconstruction across multiple heterogeneous datasets with differing channel characteristics. Suppose we collect \(M\) heterogeneous datasets:
\begin{equation}
\mathcal{D} \;=\; \{\mathcal{D}_{1},\,\mathcal{D}_{2},\,\ldots,\,\mathcal{D}_{M}\},
\end{equation}
where each dataset \(\mathcal{D}_{m}\) corresponds to a different deployment scenario (e.g., different carrier frequency, communication environment, user distribution, antenna topology). Denote the \(i\)-th sample in \(\mathcal{D}_{m}\) by \(\mathbf{H}^i_{\mathcal{D}_{m}}\), where the BS serves \(K_{m}\) users.

The unified network must be capable of processing and reconstructing CSI from all heterogeneous configurations. Accordingly, we formulate the overall optimization objective as
\begin{subequations}
\begin{align}
\min_{\boldsymbol{\theta}_{\mathcal{E}},\,\boldsymbol{\theta}_{\mathcal{R}}} &
\sum_{m=1}^{M} \sum_{i=1}^{|\mathcal{D}_{m}|} \, 
 \mathcal{L}\Bigl(
\hat{\mathbf{H}}^i_{\mathcal{D}_{m}} ,\;\mathbf{H}^i_{\mathcal{D}_{m}}\Bigr),    \\
\mbox{s.t.}\quad &
\hat{\mathbf{H}}^i_{\mathcal{D}_{m}} = \mathcal{R}\bigl(\bigl\{      \mathcal{D}(\mathbf{s}_{k})\bigr\}_{k=1}^{K_m};\,\boldsymbol{\theta}_{\mathcal{R}}\bigr),  \\
& \quad  s_k = \mathcal{Q}\bigl(\mathcal{E}_k(\mathbf{H}^i_{\mathcal{D}_{m}, \,k};\,\boldsymbol{\theta}_{\mathcal{E},\, k})\bigr),  ~~\forall k
\end{align}
\end{subequations}
where \(\mathcal{L}(\hat{\mathbf{H}},\,\mathbf{H})\) is the evaluation function, which is typically chosen as the normalized mean squared error (NMSE). 

\section{Wireless Foundation Model for Heterogeneous CSI Feedback}

\begin{figure*}[htbp]
    \centering
    \vspace{-1em}
    \includegraphics[width=1\linewidth]{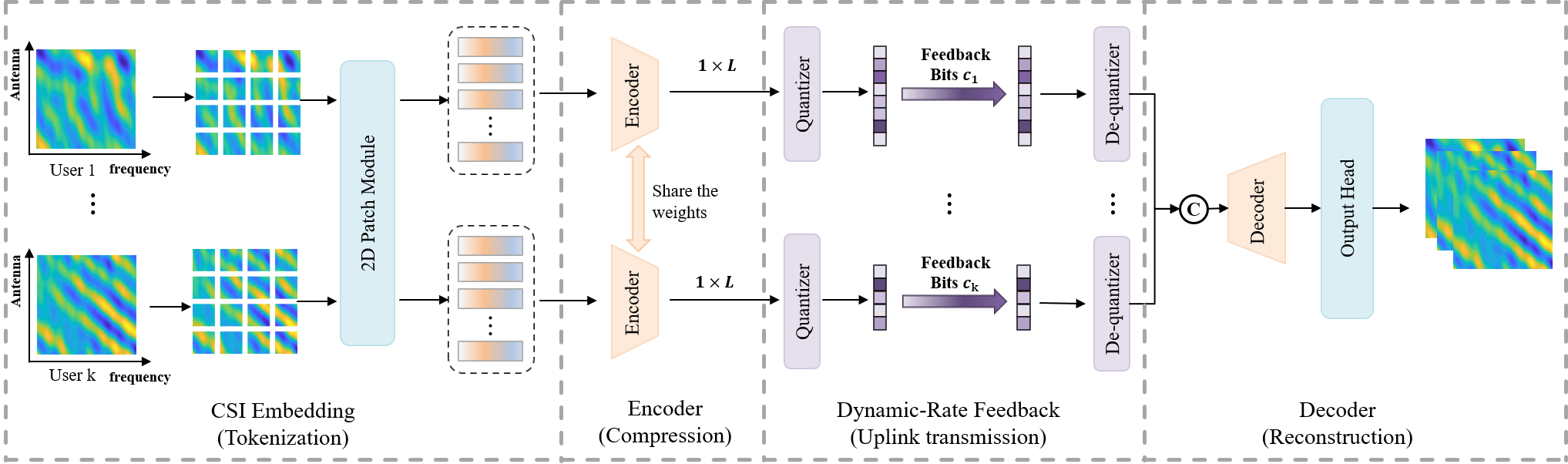}
    \caption{The proposed WiFo-CF is composed of four main modules: (i) CSI embedding module; (ii) encoder; (iii) dynamic-rate feedback module; (iv) decoder.}
    \label{network}
    \vspace{-1em}
    \end{figure*}

Although heterogeneous CSI datasets exhibit diverse characteristics in terms of data format and statistical distribution, they are fundamentally wireless channel data that conform to physical propagation laws, and thus exhibit inherent correlation that can be mined and leveraged. 
Joint training on such heterogeneous datasets can therefore significantly enhance a model’s universal channel representation capability and generalization performance compared to training on a single dataset. To fully exploit this advantage, we propose a heterogeneous CSI feedback scheme based on a wireless foundation model, termed WiFo-CF. The proposed scheme employs a carefully designed network architecture to accommodate channel data inputs of varying dimensions and devises a corresponding pre-training strategy to endow the model with robust generalization and channel-representation capabilities. The network components are illustrated in Fig. \ref{network}. In the following sections, we provide a detailed description of the WiFo-CF network components and the associated training procedure.

\subsection{Network Structure}

A multi-user scalable autoencoder (MUAE) architecture is proposed to support the multi-user CSI feedback scenario considered in this work. Specifically, identical encoder networks are deployed at all UEs, enabling flexible adaptation to varying numbers of users. Upon obtaining the estimated downlink CSI, each user independently compresses its channel and transmits the codeword to the BS via the uplink. A unified decoder at the BS jointly processes the feedback from all users and reconstructs their CSI, effectively exploiting the underlying inter-user channel correlations. As illustrated in Fig. \ref{network}, the proposed architecture comprises four modules: CSI Embedding, Encoder, Dynamic-Rate Feedback, and Decoder. Each module is described in detail in the following subsections.

\subsubsection{CSI Embedding}

Each user's operation can be regarded as an independent parallel process. Therefore, we describe the model flow for a representative user $k$. Given the estimated downlink channel matrix $\mathbf{H}_k \in \mathbb{C}^{N_t \times N_c}$, the first step is to convert it into a real-valued tensor $\overline{\mathbf{H}}_k \in \mathbb{R}^{2 \times N_t \times N_c}$, where the first and second channels represent the real and imaginary parts of $\mathbf{H}_k$, respectively.

To make the input compatible with the Transformer module, the channel matrix is then partitioned into patches using a non-overlapping 2D convolution operation~\cite{dosovitskiy2020image}. The convolution kernel size $(p_n, p_f)$ defines the patch size along the spatial and frequency dimensions, respectively. This process can be formulated as:
\begin{equation}
    \mathbf{H}_{\text{conv}} = \text{Conv2d}(\overline{\mathbf{H}}_k),
\end{equation}
where $\text{Conv2d}(\cdot)$ denotes a 2D convolution with 2 input channels and $d_{\text{enc}}$ output channels. Then, the output $\mathbf{H}_{\text{conv}} \in \mathbb{R}^{d_{enc} \times \frac{N_t}{p_n} \times \frac{N_c}{p_f}}$ is flattened and used as the CSI tokens $\mathbf{H}_{\text{emb}} \in \mathbb{R}^{L \times d_{enc}}$.

\subsubsection{Encoder Design}

\begin{figure}[htbp]
    \centering
    \includegraphics[width=1\linewidth]{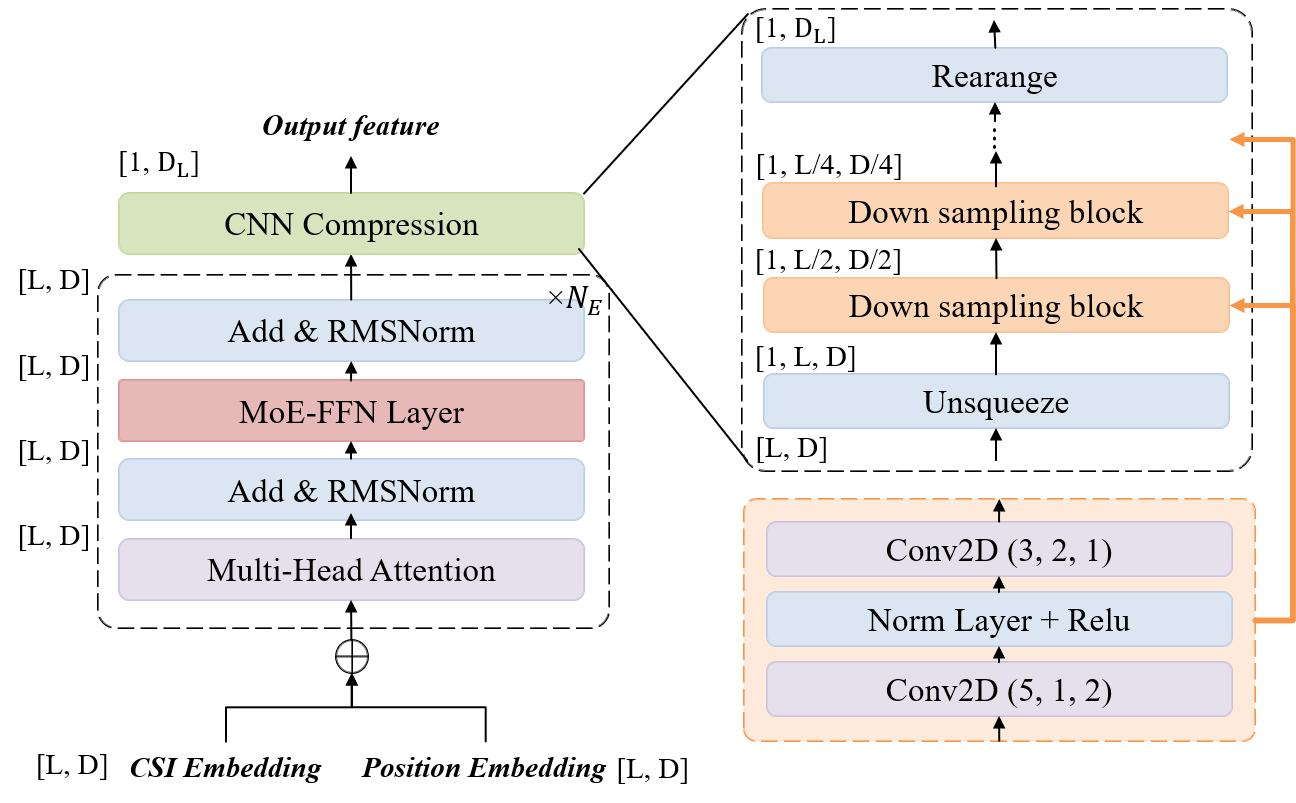}
    \caption{An illustration of the encoder architecture in WiFo-CF.}
    \label{encoder}
    \vspace{-1em}
    \end{figure} 
    
The encoder module is responsible for extracting and compressing features from the input CSI tokens. As illustrated in Fig.~\ref{encoder}, it consists of three components: positional encoding, the Transformer blocks for feature extraction, and a downsampling module for feature compression.

For positional encoding, we adopt the standard absolute positional encoding scheme as follows:
\begin{equation}
    \mathbf{H}_{enc}^{0} =  \mathbf{H}_{enc} + \mathbf{P}_{absolute},
\end{equation}
where ${P}_{absolute} \in \mathbb{R}^{L \times d_{enc}}$ denotes the positional encoding, and
\begin{subequations}
\begin{align}
\bm{P}_{absolute}(2i,j)=\sin (\frac{j}{10000^{\frac{2i}{d_{enc}}}}), \label{PE0} \\
\bm{P}_{absolute}(2i+1,j)=\cos (\frac{j}{10000^{\frac{2i}{d_{enc}}}}). \label{PE1} 
\end{align}
\end{subequations}
 
Subsequently, a series of stacked Transformer blocks \cite{vaswani2017attention} is employed to extract features from the input tokens. To improve training stability, RMSNorm \cite{zhang2019root} is applied to normalize the input features. In addition, to enable the adaptive processing of diverse CSI patches, the feed-forward layers in the Transformer blocks are replaced with a Mixture-of-Experts (MoE) structure.
\begin{subequations}
\begin{align}
 \mathbf{U}_{enc}^{l}=\text{RMSNorm}(\text{MHA}(\mathbf{H}_{enc}^{l-1}) + \mathbf{H}_{enc}^{l-1}), \label{TSA} \\
 \mathbf{H}_{enc}^{l}= \text{RMSNorm}(\text{MoE}(\mathbf{U}_{enc}^{l})+\mathbf{U}_{enc}^{l}). \label{TMOE} 
\end{align}
\end{subequations}
Here, MHA and MoE denote the multi-head attention and MoE-FFN layer, respectively. $\mathbf{H}_{enc}^{l}$ and $\mathbf{U}_{enc}^{l}$ denote the output features and hidden features of the $l$-th layer transformer, respectively. The MoE-FFN module consists of $N$ expert networks, each sharing the same architecture as a standard FFN \cite{liu2024deepseek}. For each input CSI token $x \in \mathbb{R}^{d_{enc}}$, a gating network $G(\cdot)$ produces a weight vector $G(x) \in \mathbb{R}^{N}$ over all experts. A top-$k$ selection strategy is then applied to sparsely activate the most relevant experts. The MoE-FFN output can be formulated as:
\begin{equation}
\text{MoE}(x) = \sum_{i \in \mathcal{T}_k(x)} G(x)[i] \cdot \text{E}_i(x),
\end{equation}
where $\mathcal{T}_k(x)$ denotes the indices of the top-$k$ experts selected by the gating network for input $x$, and $\text{E}_i(\cdot)$ represents the $i$-th expert network.
 
Finally, a CNN-based downsampling module is applied to the output features of the Transformer blocks to obtain the compressed feature map. This design ensures size-independence and enables robust processing of CSI with varying shapes. The compressed feature map is then flattened to obtain a low-dimensional feature vector $s_k \in \mathbb{R}^{D_L}$.

\subsubsection{Dynamic-Rate Feedback}

The output of the neural network must be quantized into a bitstream before being transmitted to the BS via the uplink. Correspondingly, upon receiving the bitstreams from each user, the BS first performs dequantization to obtain the corresponding feature vector $\hat{s}_k$ for each user, then feeds the reconstructed vectors into the decoder network, just as shown in Fig. \ref{network}.

\textcolor{black}{
In this work, we adopt a non-uniform $\mu$-law quantizer, which allocates higher resolution to low-amplitude signals. Its low complexity and adaptability to heterogeneous channel conditions make it a practical solution, as also demonstrated in prior work \cite{guo2020convolutional}.} Given an input $x \in [-1, 1]$, the $\mu$-law compression function is defined as:
\begin{equation}
F(x) = \text{sgn}(x) \cdot \frac{\ln(1 + \mu |x|)}{\ln(1 + \mu)},
\end{equation}
where $\mu > 0$ is a hyperparameter controlling the degree of non-linearity, and $\text{sgn}(x)$ denotes the sign function.

Let $b_k$ denote the quantization bit-width for user $k$. For a feature vector of length $D_L$, the resulting bitstream has a total length of $B_k = D_L \cdot b_k$.

In practical systems, users often have heterogeneous bandwidth constraints and transmission priorities. To reflect this, we allow each user to employ a quantizer with a different bit-width. \textcolor{black}{These bit-width configurations are assumed to be known at the BS, either through pre-configuration or via lightweight signaling, to ensure accurate CSI reconstruction.}
This setting not only aligns better with real-world system requirements but also brings benefits from the perspective of joint reconstruction at the BS. Since the decoder jointly reconstructs the CSI of all users, the channel correlations from high-precision feedback users can help compensate for the low-precision ones. As a result, the reconstruction accuracy for all users is improved, effectively enhancing the lower bound of system-level communication efficiency, particularly for users constrained by limited uplink bandwidth.

\subsubsection{Decoder Design}

\begin{figure*}[htbp]
    \centering
    \includegraphics[width=0.9\linewidth]{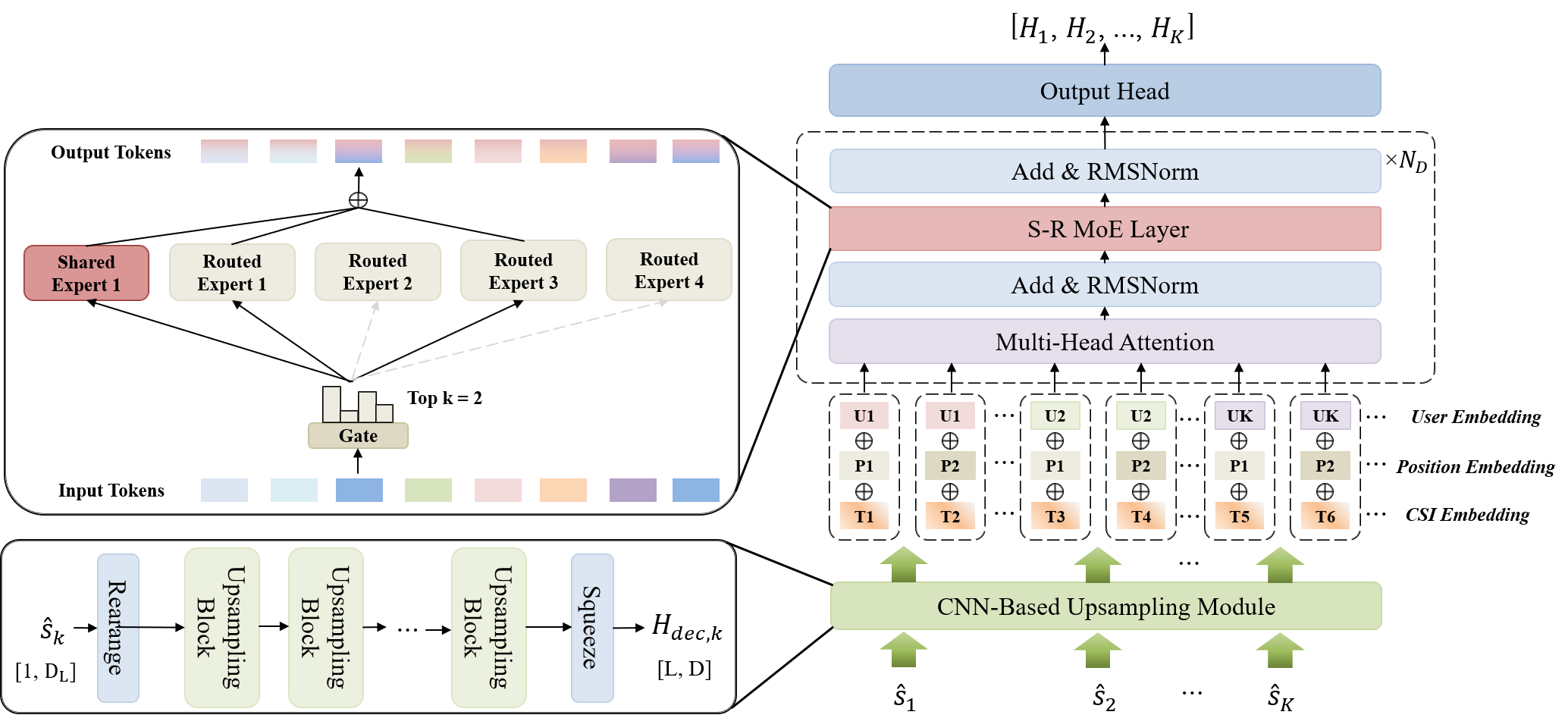}
    \caption{An illustration of the decoder architecture in WiFo-CF.}
    \label{decoder}
    \vspace{-1em}
    \end{figure*} 

\textcolor{black}{
The decoder module, deployed at the BS, jointly reconstructs the full CSI of all users from their received codewords. As illustrated in Fig.~\ref{decoder}, it comprises a CNN-based up-sampling module, a positional encoding module, several Transformer blocks, and an output head.}

\textcolor{black}{
The up-sampling module symmetrically mirrors the down-sampling design, mapping each user's codewords to a complete token sequence $\mathbf{H}_{\mathrm{dec},k}\in\mathbb{R}^{L\times d_{\mathrm{dec}}}$. To enable joint multi-user decoding, we incorporate positional encodings that embed both user identity and token position, allowing the model to capture inter- and intra-user correlations effectively.} Specifically,
\begin{subequations}
\begin{align}
\mathbf{H}^{0}_{\mathrm{dec},k} &= \mathbf{H}_{\mathrm{dec},k} + \mathbf{P}_{\mathrm{absolute}} + \mathbf{U}_{k},\\
\mathbf{U}_{k} &= \mathbf{J}_{L}\,\mathbf{U}_{\mathrm{absolute}}[k],
\end{align}
\end{subequations}
where 
\(\mathbf{H}^{0}_{\mathrm{dec},k} \in \mathbb{R}^{L \times d_{\mathrm{enc}}}\) denotes the encoded token sequence for the \(k\)-th user,
\(\mathbf{P}_{\mathrm{absolute}}\) and \(\mathbf{U}_{\mathrm{absolute}}\) follow the same standard absolute positional encoding used in the encoder.
\(\mathbf{J}_{L}\in\mathbb{R}^{L\times1}\) is an all‑ones column vector, and the multiplication follows a broadcasting mechanism along the token dimension to replicate \(\mathbf{U}_{\mathrm{absolute}}[k]\) across all \(L\) tokens.
 
We then concatenate the encoded token sequences of all users along the sequence dimension and feed the result into the Transformer blocks. Formally, the input to the Transformer is given by:
\begin{equation}
\resizebox{0.88\hsize}{!}{$
\begin{aligned}
\mathbf{H}^{0}_{\mathrm{dec}} = \left[ \mathbf{H}^{0}_{\mathrm{dec},1} \;\middle|\; \mathbf{H}^{0}_{\mathrm{dec},2} \;\middle|\; \cdots \;\middle|\; \mathbf{H}^{0}_{\mathrm{dec},K} \right] \in \mathbb{R}^{(K \cdot L) \times d_{\mathrm{enc}}},
\end{aligned}$}
\end{equation}
where \([\cdot\,|\,\cdot]\) represents concatenation along the sequence dimension.

This unified representation is subsequently fed into the Transformer decoder, which exploits its strong modeling capacity to jointly reconstruct the CSIs of all users. To better capture the intricate inter-user dependencies, we adopt an asymmetric encoder-decoder architecture, in which the decoder is assigned a higher feature dimensionality and greater depth than the encoder. This design aligns with the practical assumption that the BS possesses substantially more computational resources than UE, facilitating effective modeling of complex inter-user interactions and accurate reconstruction of high-dimensional, multi-user CSI.

In addition, to further enhance the Transformer’s ability to model multi-user CSI, we replace the standard FFN modules with a Mixture of Shared and Routed Experts (S-R MoE) architecture.
The S-R MoE module comprises \(N_s\) Shared Experts and \(N_r\) Routed Experts, all sharing the same architecture as a standard FFN \cite{liu2024deepseek}. For each input CSI token \(x \in \mathbb{R}^{d_{\mathrm{dec}}}\), a gating network \(G(\cdot)\) produces a weight vector \(G(x) \in \mathbb{R}^{N_r}\), and a top-\(k\) sparse selection strategy is applied to activate the most relevant Routed Experts. The output of the module is given by:
\begin{equation}
\resizebox{0.88\hsize}{!}{$
\begin{aligned}
\text{S-R\,MoE}(x) = \sum_{j=1}^{N_s} \mathrm{S\text{-}E}_j(x) + \sum_{i \in \mathcal{T}_k(x)} G(x)[i] \cdot \mathrm{R\text{-}E}_i(x),
\end{aligned}$}
\end{equation}
where \(\mathrm{S\text{-}E}_j(\cdot)\) and \(\mathrm{R\text{-}E}_i(\cdot)\) represent the \(j\)-th Shared Expert and the \(i\)-th Routed Expert, respectively, and \(\mathcal{T}_k(x)\) denotes the indices of the top-\(k\) Routed Experts selected by the gating network for input \(x\).

In this design, Shared Experts are used to capture representations that are common across users, enabling the model to exploit shared structural priors in the CSI. In contrast, Routed Experts are dedicated to learning user-specific variations, allowing for personalized adaptation to each user's channel characteristics. This expert separation facilitates a more efficient modeling of both shared and distinct features across users, which is particularly beneficial in multi-user scenarios with heterogeneous CSI distributions. Moreover, unlike conventional approaches that require retraining or zero-padding to accommodate varying numbers of users, the S-R MoE architecture can naturally handle an arbitrary number of users, thereby improving the model’s generalization capability while reducing both the retraining overhead and the computational inefficiency associated with input padding.

Finally, we adopt a unified output head to reconstruct CSI data from heterogeneous sources, in contrast to existing methods that are confined to a single simulated data source. This design enhances WiFo‑CF's flexibility by enabling it to learn from diverse data patterns and distributions. During training, the overall loss is computed by aggregating the reconstruction losses of CSIs from different generative sources (see Section~\ref{pre-train}), which further promotes robust generalization across data domains.

\subsection{Pre-training Dataset}


 \begin{table*}[t]
\centering
\renewcommand\arraystretch{1.3}  
\caption{Overview of Pre-training and Testing Datasets of LH-CDF.}
\label{tab: Dataset}
\begin{tabular}{ccccccccccc}
\toprule
Set Type & Data Gen. Source & Dataset ID & Fre. (GHz) & BS. Ant. & UE. Ant. & Num Subc. & UE Speed (km/h) & Samples (k) \\ 
\midrule
\multirow{3}{*}{\makecell[c]{Pre-training Set \\ (full shot)}} &  QuaDriGa & Q1-Q8 & 1.0-6.0 & 32-64 & 1-4 & 16-64 & 0-100 &  480 \\  
& DeepMIMO & D1-D2 & 3.4-3.5 & 16-32 & 1-2 & 32 & N/A &  120  \\  
& Argos & A1-A6 & 2.4 & 64 & 1 & 32 & 0-30 &  90  \\  
\midrule
\multirow{5}{*}{\makecell[c]{Testing Set \\ (zero shot)}} &  QuaDriGa & Q9-Q16 & 1.0-28.0 & 32-256 & 1-4 & 16-64 & 0-300 &  480 \\  
& DeepMIMO & D3-D4 & 28.0-60.0 & 16-32 & 1-2 & 64 & N/A &  120  \\  
& SynthSoM & S1-S2 & 28.0 & 64 & 1-2 & 64 & N/A &  10  \\  
& Argos & A7-A8 & 2.4 & 64 & 1 & 32 & 0-30 &  30  \\  
& Dichasus & H1-H2 & 1.27-3.44 & 32 & 1 & 32 & N/A &  30  \\ 
\bottomrule
\end{tabular}
\end{table*}

Training foundation models requires large-scale, high-quality datasets. LWM utilizes ray-tracing-based channel data generated from multiple scenarios using DeepMIMO, while LAM constructs statistically modeled channel datasets with varying user distributions using QuaDriGa. However, both approaches rely solely on a single data generation paradigm, which may lead the model to overfit specific synthetic patterns and hinder its generalization capability. 

To address this issue, we build the first large-scale, multi-level heterogeneous channel dataset for CSI feedback (LH-CDF). 
\textcolor{black}{In contrast to prior CSI datasets, LH-CDF integrates multiple data-generation pipelines and covers a wide range of channel configurations (antenna arrays, frequency bands, user distributions, and cell scenarios) across varied system setups. The resulting heterogeneity mirrors practical deployment diversity. In addition, synchronized multi-user CSI snapshots enable studies of spatial correlation and joint-feedback schemes. We next present the details of LH-CDF, organized by data-generation methodology.}
\begin{itemize}
\item \textbf{Statistical Modeling-based:}
We generate a diverse set of statistically modeled CSI samples using QuaDriGa \cite{jaeckel2014quadriga}, comprising 16 sub-datasets. The covered scenarios include UMa, UMi, RMa, and Indoor, under both LoS and NLoS conditions. Detailed parameter configurations are provided in Table~\ref{tab: Dataset}.
Furthermore, each sub-dataset is further divided into four secondary sub-datasets, sharing the same communication configurations but differing in user distribution—consistent with the dataset construction strategy used in LAM.

\item \textbf{Ray-tracing-based:}
We incorporate both DeepMIMO \cite{alkhateeb2019deepmimo} and SynthSoM \cite{cheng2025synthsom} datasets to obtain ray-tracing-based CSI data. The DeepMIMO dataset, built upon Wireless InSite, provides high-fidelity ray-tracing data across four sub-datasets corresponding to the O1 scenario. Parameter settings are detailed in Table~\ref{tab: Dataset}.
The SynthSoM dataset constructs a highly realistic scenario with a dense deployment of dynamic scatterers such as pedestrians. CSI data is generated using Sionna RT \cite{hoydis2022sionna}, resulting in two sub-datasets based on the PKU campus scenario. The specific configuration details are listed in Table~\ref{tab: Dataset}.
Similar to the statistically modeled part, each sub-dataset here is also divided into four secondary sub-datasets to separately reflect intra-dataset heterogeneity (user distribution) and inter-dataset heterogeneity (CSI dimension, scene, and generation method).

\item \textbf{Real-world Measured-based:}
We further incorporate real-world measurements by collecting and organizing eight datasets from Argos \cite{shepard2016understanding} and two from Dichasus \cite{dichasus2021}. These datasets cover a broad range of environments, including indoor classrooms, large factories, outdoor LoS, and outdoor NLoS scenarios. Detailed parameter configurations are provided in Table~\ref{tab: Dataset}.
\end{itemize}

In summary, {LH-CDF comprises over 1 million samples} and, to the best of our knowledge, represents the {largest dataset designed for CSI feedback tasks}, and the {first to simultaneously integrate three distinct channel generation methodologies}. By pre-training on such a diverse and heterogeneous dataset, {WiFo-CF} is empowered to learn more generalizable channel representations, thereby enhancing its adaptability to various heterogeneous deployment scenarios.

\subsection{Model Training} \label{pre-train}

\begin{figure}[htbp]
    \centering
    \includegraphics[width=1\linewidth]{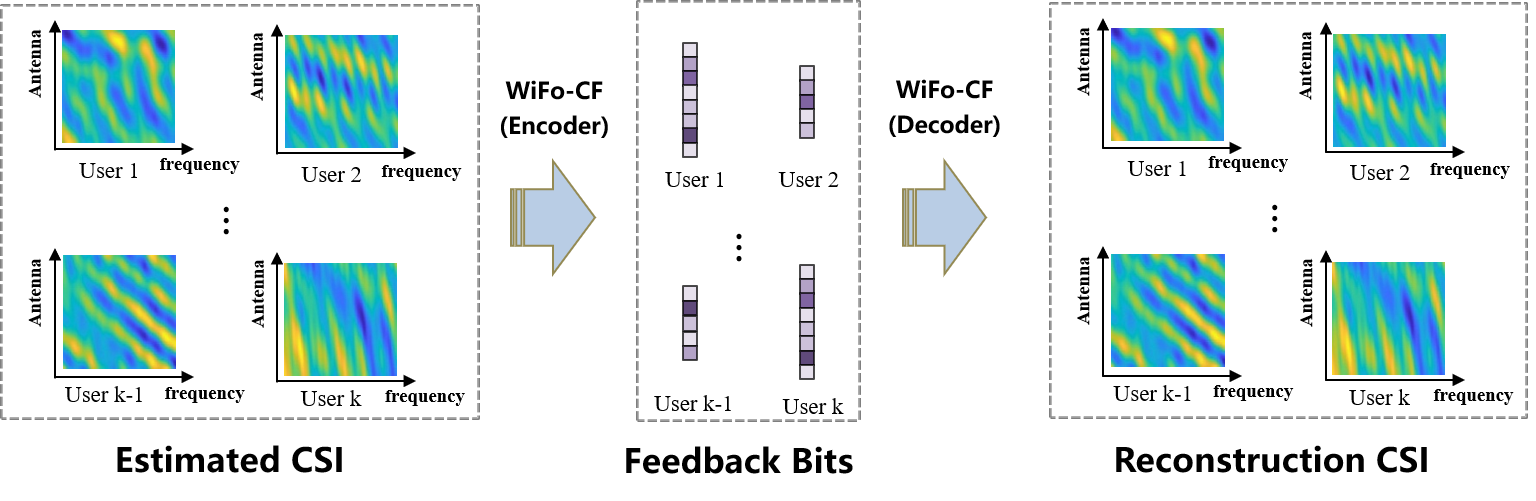}
    \caption{An illustration of the designed pre-training tasks.}
    \label{pre-training task}
    \end{figure}

The CSI feedback task exhibits a natural self-supervised learning property, making it particularly suitable for large-scale self-supervised pre-training. In contrast to existing CSI feedback methods that are typically trained and evaluated on a single dataset with fixed configurations, {WiFo-CF} is pre-trained on a large-scale heterogeneous dataset to learn more generalizable channel representations. This enables the model to effectively generalize across diverse deployment scenarios and configuration settings.
Furthermore, to equip {WiFo-CF} with the capability of handling {multi-user} and {multi-rate} CSI compression and reconstruction, we design a set of specialized pre-training tasks, as illustrated in Fig.~\ref{pre-training task}. The overall pre-training procedure is summarized in Algorithm~\ref{alg:pre-train}, which outlines the detailed steps for optimizing WiFo-CF under the proposed self-supervised learning (SSL) framework.
During pre-training, the feedback bitrate for each user is randomly sampled, enabling WiFo-CF to adapt to various bandwidth constraints. At the same time, the model is trained to jointly decode the CSI of multiple users, allowing it to exploit inter-user channel correlations and improve reconstruction accuracy.

The training objective is to minimize the reconstruction loss, with the mean squared error (MSE) adopted as the loss function:
\begin{equation} \label{mseloss}
    \mathcal{L}_{\text{rec}} = \frac{1}{K_m} \sum_{k=1}^{K_m} \left\| \hat{\mathbf{H}}^i_{\mathcal{D}_m, k} - \mathbf{H}^i_{\mathcal{D}_m, k} \right\|^2.
\end{equation}

To prevent routing collapse and promote balanced expert utilization, we further introduce a load balancing loss, inspired by the formulation in~\cite{fedus2022switch}. The load balancing loss is computed as the scaled dot-product between the token distribution vector $\mathbf{f}$ and the routing probability vector $\mathbf{P}$:
\begin{equation}  \label{auxloss}
    \mathcal{L}_{\text{lb}} =  N \cdot \sum_{i=1}^{N} f_i \cdot P_i,
\end{equation}
\textcolor{black}{where $f_i = \frac{1}{T} \sum_{x \in \mathcal{B}} \mathbb{I} \left\{ \arg\max p(x) = i \right\}$ denotes the fraction of tokens routed to expert $i$ with $\mathbb{I}\left\{ \cdot \right\}$ being the indicator function and $P_i = \frac{1}{T} \sum_{x \in \mathcal{B}} p_i(x)$ is the average router probability allocated to expert $i$.}

The total loss is the weighted sum of the reconstruction loss and the auxiliary loss:
\begin{equation}
\mathcal{L} = \mathcal{L}_{\text{rec}} + \beta_1 \cdot \mathcal{L}_{\text{lb}},
\end{equation}
where $\beta_1$ is a hyperparameter that controls the strength of the auxiliary loss, typically set to a small value such as $0.01$.

\begin{algorithm}[!ht]
    \renewcommand{\algorithmicrequire}{\textbf{Input:}}
    \renewcommand{\algorithmicensure}{\textbf{Output:}}
    \caption{Self-Supervised Pre-Training of WiFo-CF}
    \label{alg:pre-train}
    \begin{algorithmic}[1]
        \REQUIRE Pre-training datasets $\mathcal{D} = \{D_1, D_2, \ldots, D_L\}$; quantization bit width range $B_r$; user quantity range $K_r$; number of training epochs $Epoch$; balancing factor $\beta_1$
        \ENSURE Encoder parameters $\boldsymbol{\Theta}_e$; decoder parameters $\boldsymbol{\Theta}_d$
        
        \STATE Initialize $\boldsymbol{\Theta}_e = \boldsymbol{\Theta}_{e,0}$, $\boldsymbol{\Theta}_d = \boldsymbol{\Theta}_{d,0}$
        \FOR{$epoch = 1$ to $Epoch$}
            \STATE $Loss \leftarrow 0$
            \FOR{$i = 1$ to $L$} 
                \STATE Randomly sample quantization bit width $b$ from $B_r$, and number of users $K_m$ from $K_r$
                \STATE Divide dataset $D_i$ into batches $\{\mathcal{M}_j\}_{j=1}^{N_i}$ with $K_m$ samples per batch
                \FOR{$j = 1$ to $N_i$}
                    \STATE Encode and quantize each user's CSI: 
                    \[
                    s_k = \mathcal{Q}\big(\mathcal{E}(\mathbf{H}^i_{\mathcal{M}_j, k};\, \boldsymbol{\Theta}_e),\, b\big), \quad \forall k \in \{1,\dots,K_m\}
                    \]
                    \STATE Jointly decode reconstructed multi-user CSIs: 
                    \[
                    \hat{\mathbf{H}}^i_{\mathcal{M}_j} = \mathcal{R}\left(\left\{\mathcal{D}(s_k)\right\}_{k=1}^{K_m};\, \boldsymbol{\Theta}_d\right)
                    \]
                    \STATE Compute reconstruction loss: $\mathcal{L}_{\text{rec}}$ using (\ref{mseloss})
                    \STATE Compute expert balancing loss: $\mathcal{L}_{\text{aux}}$ using (\ref{auxloss})
                    \STATE Compute total loss: $\mathcal{L}_{\text{total}} = \mathcal{L}_{\text{rec}} + \beta_1 \cdot \mathcal{L}_{\text{aux}}$
                    \STATE $Loss \leftarrow Loss + \mathcal{L}_{\text{total}}$
                \ENDFOR
            \ENDFOR
            \STATE update $\boldsymbol{\Theta}_e$, $\boldsymbol{\Theta}_d$ by minimizing total loss; 
        \ENDFOR
    \end{algorithmic}
\end{algorithm}


\section{Experiments}

In this section, we first present the simulation settings and then evaluate the performance of the proposed WiFo-CF method from multiple perspectives, including in-distribution performance, out-of-distribution generalization, and channel representation capability for downstream channel-related tasks. We further perform a detailed hyperparameter analysis to investigate the impact of model size, dataset scale, and patch size on overall performance. Additionally, comprehensive ablation studies are conducted to validate the effectiveness of each component within the proposed framework.

\subsection{Simulation Setup}
\subsubsection{Baselines}
To validate the superiority of the proposed scheme, several model-based and deep learning-based methods are implemented as baselines.
\begin{itemize}

\item \textbf{TAVL3 \cite{li2009user}:} 
TVAL3 is a classical baseline method for CSI feedback, based on total variation minimization using an augmented Lagrangian and alternating direction algorithm. It has demonstrated superior performance compared to several other compressive sensing (CS)-based algorithms for CSI reconstruction, owing to its robustness and computational efficiency.
\item \textbf{CsiNet~\cite{wen2018deep}:}
CsiNet is a deep learning-based CSI feedback scheme based on an autoencoder architecture. We re-implemented the network following the specifications in~\cite{wen2018deep} for comparison. As it is designed for a fixed system configuration, we trained it separately on each secondary sub-dataset within the LH-CDF.

\item \textbf{CRNet~\cite{lu2020multi}:}
CRNet improves reconstruction performance by extracting multi-resolution CSI features and adopts a cosine annealing learning rate schedule with a warm-up phase to enhance training. Similar to CsiNet, it is limited to fixed system configurations; thus, we trained it independently on each secondary sub-dataset within the LH-CDF.

\item \textbf{LAM~\cite{guo2025prompt}:}
LAM is a recently proposed CSI feedback method that improves model generalization through large-scale pre-training on datasets with various user distributions. However, it still struggles with inputs of heterogeneous dimensions. Therefore, we trained it separately on each sub-dataset within the LH-CDF.

\end{itemize}

\subsubsection{Network and Pre-training Settings}

 \begin{table*}[t]
\centering
\begin{threeparttable} 
\renewcommand\arraystretch{1.3}  
\caption{Network parameters of WiFo-CF with different sizes.}
\label{tab: model size}
\begin{tabular}{ccccccccccc}
\toprule
Models & Enc. depth & Enc. width & Enc. heads & Dec. depth & Dec. width & Dec. heads & Num. expert\tnote{1} & Parameters (M) \\ 
\midrule
WiFo-CF-Small &  2 & 64 & 8 & 2 & 64 & 8 & 1-1-31 &  0.10 / 1.73 \\  
WiFo-CF-Base &  2 & 128 & 8 & 4 & 128 & 8 & 1-3-31 &  1.60 / 8.38 \\  
WiFo-CF-Large & 4 & 128 & 8 & 4 & 128 & 8 & 1-7-31 &  3.89 / 14.18 \\  
\bottomrule
\end{tabular}
\begin{tablenotes}    
\footnotesize               
\item[1] In ``Num. expert", {a}-{b}-{c} indicates the numbers of shared, activated, and router experts in the S-R MoE, respectively.
\end{tablenotes}             
\end{threeparttable}
\end{table*}

We pre-train three versions of WiFo-CF with different model sizes to accommodate deployment across devices with varying resource constraints. Each version adopts the MoE architecture to ensure scalability, with detailed configurations listed in Table~\ref{tab: model size}. In addition, the 2D patch module is implemented using a kernel size of (4, 4), and the compression ratio CR is fixed at 1/32.

During pre-training, the quantization bit-width is randomly sampled within the range of 3 to 7 bits, and the number of users jointly processed per sample is set between 2 and 6. All experiments are conducted on a server equipped with four Intel Xeon Platinum 8358P CPUs, four NVIDIA RTX 4090 GPUs, and 188~GB of RAM. The training configurations are summarized in Table~\ref{tab:training_settings}. The model is pre-trained on the datasets listed in Table~\ref{tab: Dataset}. Each dataset is first converted into a suitable input format according to the procedure described in Algorithm~\ref{alg:pre-train}, then shuffled and sequentially fed into the model for parameter updates. The hyperparameter $\beta_1$ is set to the default value of 0.01.

\begin{table}[h]
\renewcommand\arraystretch{1.3}  
\caption{Hyperparameters for network training}
\label{tab:training_settings}
\centering
\scriptsize
\begin{tabular}{|c|c|}
\hline
\makebox[0.20\textwidth][c]{\textbf{Parameter}} & \makebox[0.20\textwidth][c]{\textbf{Value}} \\ \hline
Batch size & 64 \\ \hline
Pre-training epochs & 100 \\ \hline
Fine-tuning epochs & 200 \\ \hline
Optimizer & Adam ($\beta_1=0.9$, $\beta_2=0.999$) \\ \hline
Learning rate scheduler & Cosine annealing \\ \hline
Scheduler period & 100 epochs \\ \hline
Learning rate range & [$1 \times 10^{-6}$, $1 \times 10^{-5}$] \\ \hline
\end{tabular}
\end{table}

\subsubsection{Performance Metric}

In addition to the intuitive reconstruction accuracy measured by NMSE, we further evaluate the practical performance of each model within a communication system to better inform real-world deployment. Specifically, we propose a novel metric, termed {Effective Spectral Efficiency (ESE)}, which accounts for the impact of feedback delay \cite{qi2024deep} and is defined as:
\begin{equation}
\text{ESE} = R_k \times \eta , \quad \eta = 1 - \frac{B_k}{R_{u} \cdot W \cdot T_c}
\end{equation}
Here, \(R_k\) denotes the conventional spectral efficiency without considering feedback delay, which can be calculated by Eq.~\ref{Rk_OFDM}; \(B_k\) is the number of bits used for uplink feedback; \(W\) is the uplink bandwidth; \(R_u\) is the uplink transmission rate; and \(T_c\) denotes the channel coherence time. When \(T_c\) is small, the effect of feedback delay becomes significant. The term \(\eta\), referred to as the \textit{effective communication ratio}, models the degradation in practical spectral efficiency due to the time consumed by CSI feedback, under the assumption of ideal uplink conditions with maximum data rate \(R_u\).

\subsection{Performance Evaluation}
\subsubsection{In-Distribution Performance}

Table~\ref{tab:fullshot-nmse} compares the in-distribution performance of different baselines on the test set of LH-CDF's pre-training part. Due to space limitations, we only report the average performance of each baseline on the primary-level datasets. Benefiting from large-scale pre-training on the LH-CDF, WiFo-CF acquires a generalized ability for channel compression and feedback, achieving the best performance across various heterogeneous distributions. Compared with the second-best method, WiFo-CF achieves an average gain of $2.228$ dB. 
It is worth noting that LAM requires training ten separate models to handle the test cases, as it struggles to deal with heterogeneous input dimensions. Meanwhile, lightweight models such as CsiNet suffer from limited capacity, necessitating the training of 40 individual models for different scenarios (corresponding to secondary-level datasets). In contrast, WiFo-CF handles all heterogeneous configurations with a single model, significantly reducing the overhead of model management and switching in practical deployments.

\begin{table}[ht]
\centering
\setlength{\tabcolsep}{4pt}
\caption{Full-shot evaluation results on simulation data. The \textbf{boldface} denotes the highest score, while the \underline{underline} marks the second-best result. (Metric: NMSE in dB)}
\begin{tabular}{c|ccccc}
\toprule
\textbf{Dataset} & \textbf{WiFo-CF} & \textbf{TAVL3} & \textbf{LAM} & \textbf{CRNet} & \textbf{CsiNet}  \\
\midrule
Q1   & \textbf{-7.759} & -0.501            & \underline{-3.375} & -2.939   & -2.538    \\
Q2   & \textbf{-1.149} &  0.328            & \underline{-0.853} & -0.177   & -0.051     \\
Q3   & \textbf{-6.939} & -0.663            & -5.256             & \underline{-5.433} & -4.387     \\
Q4   & \textbf{-5.904} & -0.096            & -2.289             & \underline{-3.179} & -1.602    \\
Q5   & \textbf{-4.078} &  0.329            & \underline{-2.877} & -2.626   & -1.995     \\
Q6   & \textbf{-2.642} &  0.089            & -1.114             & \underline{-2.068} & -0.381     \\
Q7   & \textbf{-2.229} &  0.367            & -0.981             & \underline{-1.073} & -0.113    \\
Q8   & \textbf{-5.879} &  0.266            & \underline{-3.540} & -2.969   & -2.757    \\
D1   & \textbf{-6.271} &  0.409            & \underline{-3.280} & -2.028   & -1.501     \\
D2   & \textbf{-6.282} &  0.386            & \underline{-3.348} & -2.889   & -1.508     \\
\midrule
\rowcolor{gray!15}
\textbf{Avg.} & \textbf{-4.913} &  0.091            & \underline{-2.685} & -2.538   & -1.683    \\
\bottomrule
\end{tabular}
\label{tab:fullshot-nmse}
\end{table}

\begin{figure}[htbp]
    \centering
    \includegraphics[width=0.8\linewidth]{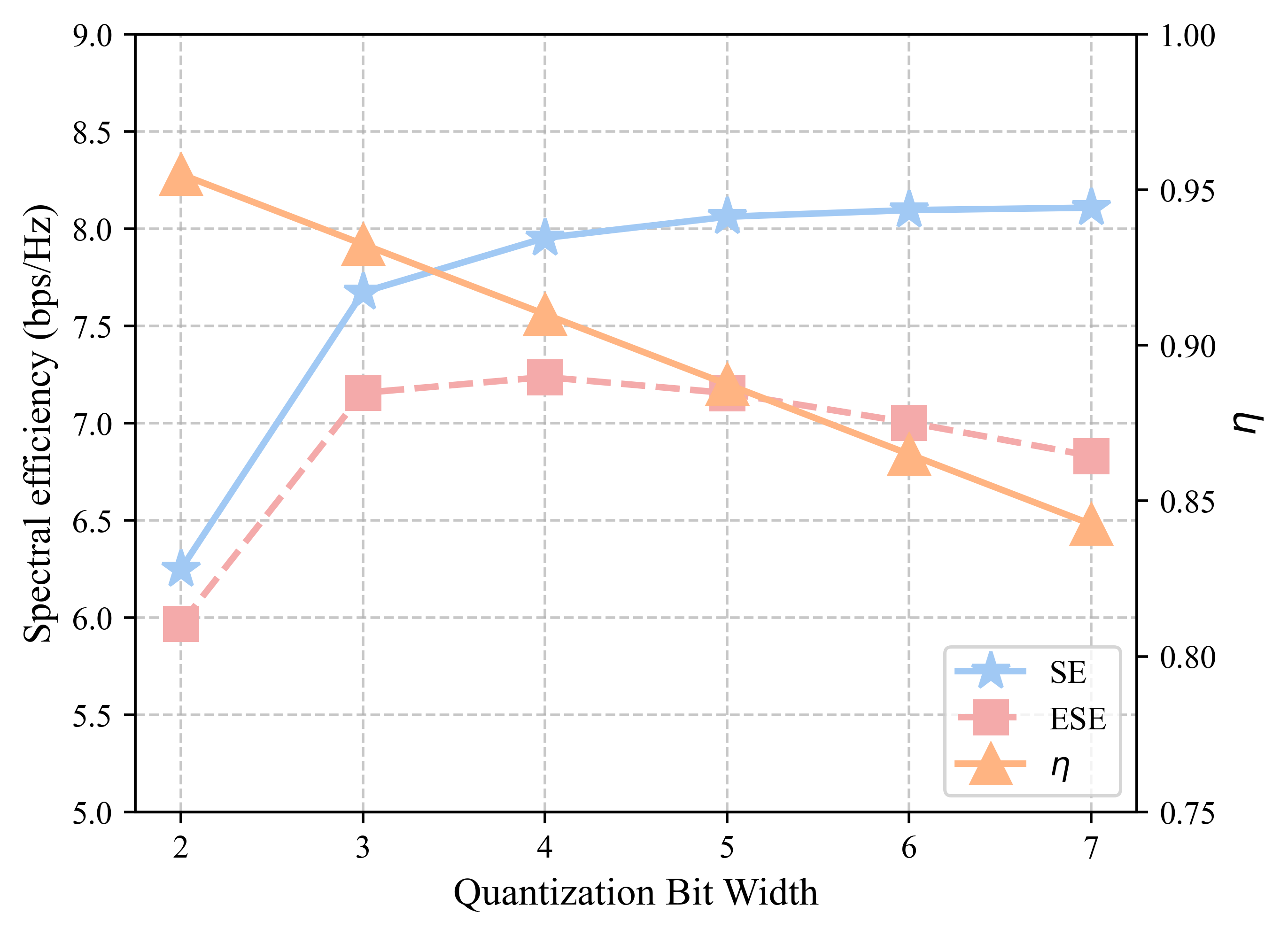}
    \caption{SE and ESE performance evaluated under various quantization bit widths.}
    \label{fig:ESE-1}
\end{figure}

Furthermore, we evaluate the multi-rate feedback capability of WiFo-CF, as illustrated in Fig.~\ref{fig:ESE-1}, where the horizontal axis denotes the quantization bit width $b$, and the vertical axes represent SE, ESE, and the effective communication ratio $\eta$, respectively. 
\textcolor{black}{
{WiFo‑CF} maintains stable operation across all tested bit widths, verifying its robustness and flexibility. 
As expected, enlarging~$b$ improves SE owing to finer quantisation. 
However, a larger~$b$ also increases the number of quantization bits for CSI feedback, thereby lengthening the uplink delay and reducing $\eta$, which ultimately offsets the performance gain and leads to a decrease in DSE beyond a certain operating point. 
Consequently, simply choosing the maximum available~$b$ is {not} always beneficial in practice.
In particular, in high-mobility or millimeter-wave scenarios where the channel coherence time is significantly shortened, it becomes critical to constrain the CSI feedback delay within the coherence interval to ensure the timeliness and validity of the acquired CSI. This consideration often favors a smaller $b$ to reduce feedback latency and maintain effective CSI acquisition. Notably, such trade-offs between CSI accuracy and acquisition delay are not adequately captured by conventional SE metrics. In contrast, the proposed ESE metric explicitly accounts for both the quantization-induced CSI quality and the latency associated with feedback, thereby providing a more faithful and practical performance measure in time-varying environments.
}

\subsubsection{Out-of-Distribution Generalization}

To evaluate the out-of-distribution generalization ability of the models, we conduct zero-shot and few-shot generalization experiments on the test set of the LH-CDF dataset. Table~\ref{tab:zeroshot-nmse} presents the zero-shot performance of WiFo-CF on the simulation datasets, along with the full-shot performance of other baseline methods. Despite encountering unseen configurations, WiFo-CF consistently achieves strong performance, outperforming most baselines that are fully trained on these datasets. On average, WiFo-CF reduces NMSE by 1.061~dB compared to the best-performing fully trained baseline, demonstrating its powerful zero-shot generalization capability.

\begin{table}[ht]
\centering
\setlength{\tabcolsep}{4pt}
\caption{Zero-shot evaluation results on simulation data. The \textbf{boldface} denotes the highest score, while the \underline{underline} marks the second-best result. (Metric: NMSE in dB)}
\begin{tabular}{c|ccccc}
\toprule
\textbf{Dataset} & \textbf{ \makecell[c]{WiFo-CF \\ (Zero Shot)}} & \textbf{TAVL3} & \textbf{LAM} & \textbf{CRNet} & \textbf{CsiNet} \\
\midrule
Q9  & \underline{-2.928} & -1.180 & \textbf{-3.153} & {-2.389} & -2.297 \\
Q10 & \textbf{-1.884} &  0.379 & -1.292 & \underline{-1.803} & -0.509 \\
Q11 & \textbf{-5.601} & -0.106 & \underline{-2.316} & {-2.186} & -1.575 \\
Q12 & \textbf{-1.445} &  0.376 & \underline{-0.990} & {-0.392} & -0.141 \\
Q13 & \textbf{-3.902} &  0.066 & -3.060 & \underline{-3.170} & -2.343 \\
Q14 & \textbf{-4.363} &  0.367 & \underline{-3.444} & {-2.945} & -2.605 \\
Q15 & \textbf{-3.973} & -0.496 & \underline{-3.044} & {-2.385} & -2.590 \\
Q16 & \textbf{-4.068} & -1.708 & \underline{-3.835} & {-2.798} & -3.042 \\
D3  & \textbf{-6.283} &  0.418 & \underline{-5.159} & {-3.882} & -3.625 \\
D4  & \textbf{-6.948} &  0.423 & \underline{-5.727} & {-5.180} & -5.377 \\
S1  & \textbf{-9.518} &  -3.256 & \underline{-7.218} & {-6.816} & -6.513 \\
\midrule
\rowcolor{gray!15}
\textbf{Avg.} & \textbf{-4.628} & -0.428 & \underline{-3.567} & {-3.086} & -2.783 \\
\bottomrule
\end{tabular}
\label{tab:zeroshot-nmse}
\end{table}

\begin{figure}[htbp]
    \centering
    \includegraphics[width=0.8\linewidth]{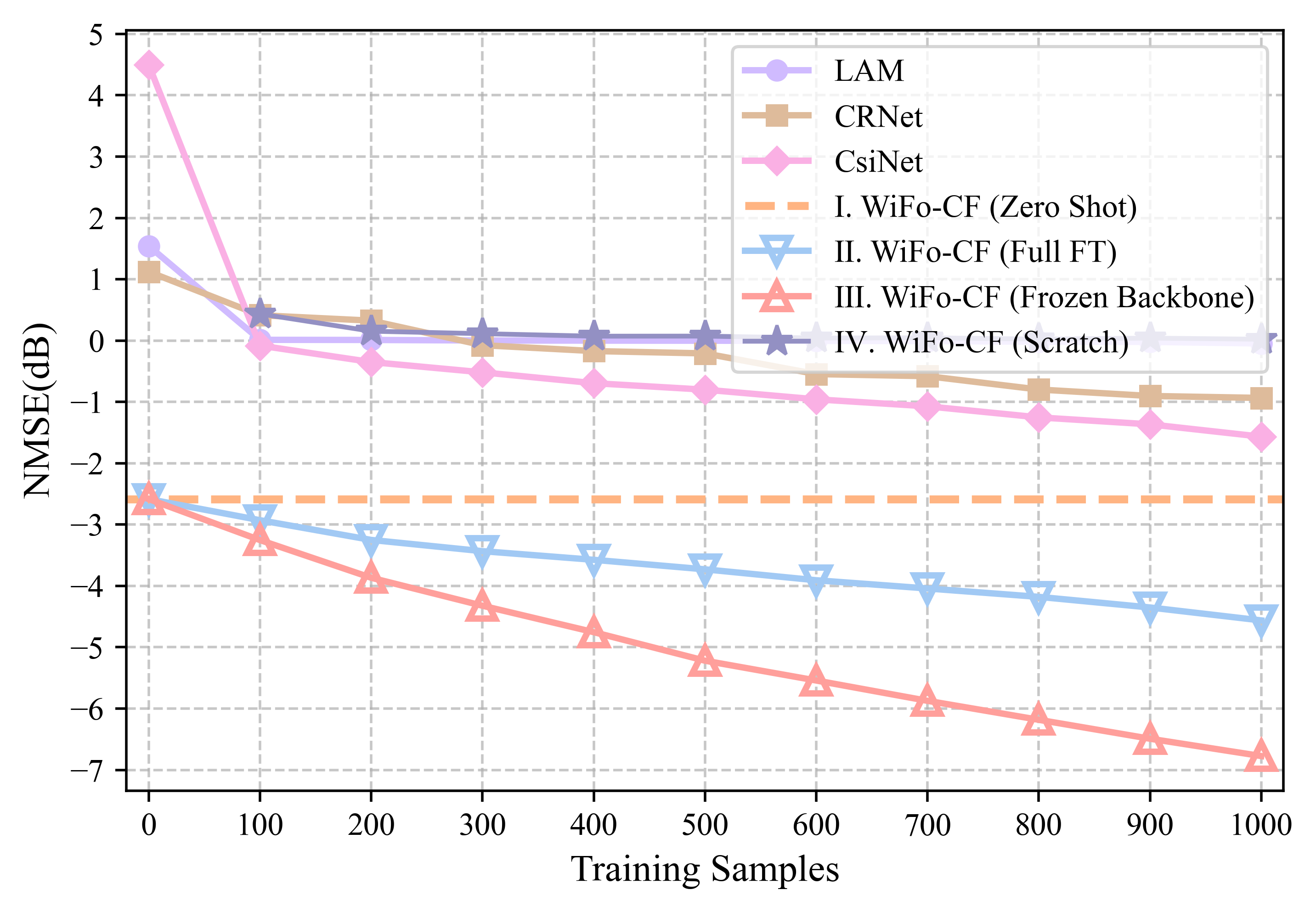}
    \caption{Few-shot Generalization Performance on the A7 Dataset.}
    \label{fig:Few shot A42}
\end{figure}

\begin{figure}[htbp]
    \centering
    \includegraphics[width=0.8\linewidth]{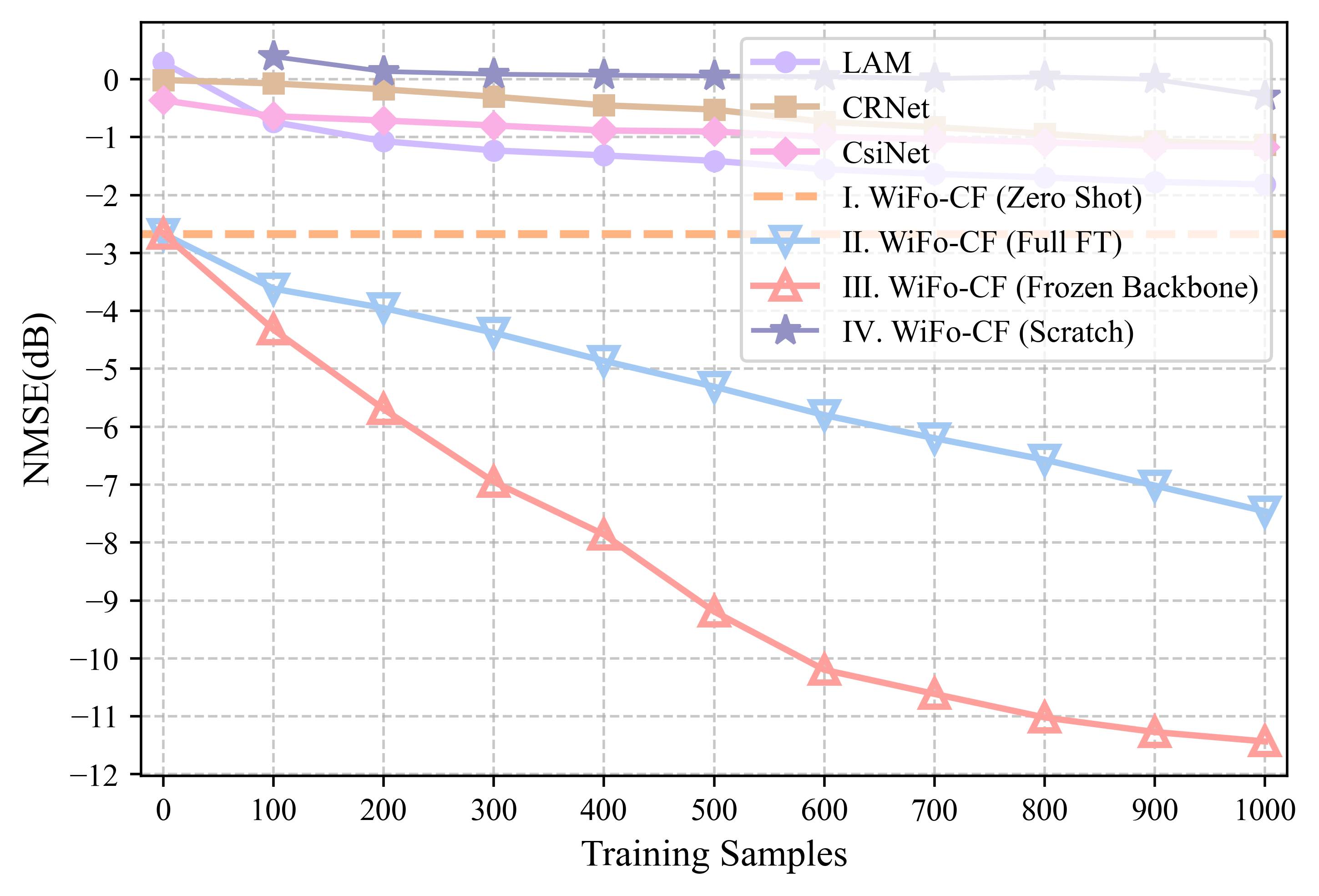}
    \caption{Few-shot Generalization Performance on the H1 Dataset.}
    \label{fig:Few shot H1}
\end{figure}

In addition, Fig.~\ref{fig:Few shot A42}-\ref{fig:Few shot H1} illustrate the few-shot performance of WiFo-CF compared with other baseline methods on various real-world datasets. As detailed in Table~\ref{tab:few shot config}, we consider four variants of WiFo-CF: 
\begin{itemize}
    \item \textbf{WiFo-CF (Zero Shot)}: directly applied to the target dataset without any fine-tuning,
    \item \textbf{WiFo-CF (Full FT)}: full-parameter fine-tuning on the target dataset,
    \item \textbf{WiFo-CF (Frozen Backbone)}: freezing the pre-trained backbone while only fine-tuning the output head,
    \item \textbf{WiFo-CF (Scratch)}: training the model from scratch without loading pre-trained weights.
\end{itemize}
All baseline methods are fully fine-tuned on their respective target datasets. The few-shot evaluation yields several important observations.
{First}, as illustrated by \textbf{WiFo-CF (Zero Shot)}, our model exhibits strong generalization even without any fine-tuning. It achieves an NMSE improvement of {3.001~dB} over the best-performing baseline under the zero-shot setting, and notably, it even surpasses all baselines trained with 1000 labeled samples. This result highlights the effectiveness of the knowledge encoded during pre-training.
{Second}, under full fine-tuning, as shown by \textbf{WiFo-CF (Full FT)}, WiFo-CF consistently outperforms all baseline methods. This can be attributed to the general channel compression knowledge learned during pre-training, which allows the model to rapidly adapt to new deployment scenarios using only limited data, without the need to relearn fundamental channel characteristics.
{Third}, we observe that \textbf{WiFo-CF (Frozen Backbone)}, which keeps the pre-trained backbone frozen and only fine-tunes the output head, surprisingly achieves even better performance than the fully fine-tuned version. This suggests that the pre-trained backbone captures highly generalizable representations of channel features, and adapting only the output layer to new CSI distributions is sufficient for effective transfer. Moreover, fine-tuning fewer parameters not only reduces training overhead but also mitigates the risk of overfitting, underscoring the practicality and adaptability of WiFo-CF in dynamic deployment environments.
{Finally}, by introducing \textbf{WiFo-CF (Scratch)} as a control setting, we observe a clear degradation in few-shot generalization ability. This further highlights the importance of the general knowledge acquired by WiFo-CF through large-scale pre-training.

\begin{table}[h]
\centering
\vspace{-1em}
\renewcommand\arraystretch{1.3}  
\caption{Comparison of different schemes with pre-training datasets and the number of network parameters.}
\begin{tabular}{|c|c|c|}
\hline
\textbf{Scheme} & \textbf{\makecell[c]{Pre-training  dataset}} & \textbf{\makecell[c]{Trainable / Total \\ params (M)}}  \\ \hline
\makecell[c]{WiFo-CF  (Zero Shot)} & LH-CDF & 0 / 8.38 \\ \hline
\makecell[c]{WiFo-CF  (Full FT)} & LH-CDF & 8.38 / 8.38 \\ \hline
\makecell[c]{WiFo-CF \\ (Frozen Backbone)} & LH-CDF & 0.64 / 8.38 \\ \hline
\makecell[c]{WiFo-CF  (Scratch)} & N/A & 8.38 / 8.38 \\ \hline
LAM & A1 & 0.54 / 0.54 \\ \hline
CsiNet & A1.1 & 0.14 / 0.14 \\ \hline
CRNet & A1.1 & 0.27 / 0.27 \\ \hline
\end{tabular}
\label{tab:few shot config}
\vspace{-1em}
\end{table}

\subsubsection{Downstream Evaluation with Indoor Localization as a Case Study}

\begin{figure*}[htbp]
    \centering
    \includegraphics[width=\linewidth]{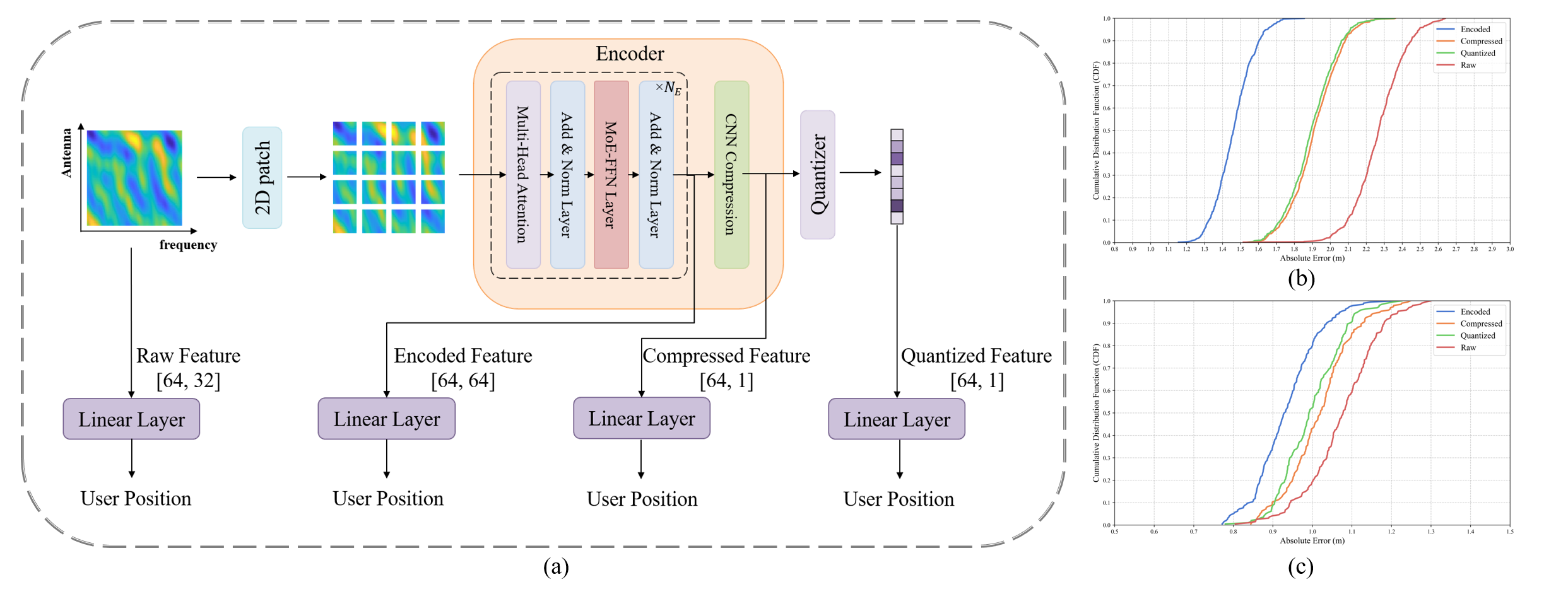}
    \vspace{-1em}
    \caption{An illustration of downstream task fine-tuning based on WiFo-CF, along with corresponding experimental results.}
    \label{fig:Downstream IP}
 
\end{figure*} 

\textcolor{black}{
As previously discussed, WiFo-CF learns generalizable channel representations through large-scale pretraining. To evaluate its cross-task transferability, we consider indoor localization as a representative downstream task, where the goal is to predict a user's 2D coordinates from the input CSI. As illustrated in Fig.~\ref{fig:Downstream IP} (a), we extract three types of features from the pretrained WiFo-CF encoder: the high-dimensional \textit{Encoded Feature} from the Transformer block, the low-dimensional \textit{Compressed Feature} from the CNN-based compression module, and the \textit{Quantized Feature} after quantization. All features are passed through identical localization heads to ensure a fair comparison.
Figs.~\ref{fig:Downstream IP} (b) and \ref{fig:Downstream IP} (c) report the average localization error for each feature type when using a single-layer and a multi-layer localization head, respectively. The \textit{Encoded Feature} consistently achieves over 10\% improvement in localization accuracy compared to raw CSI, demonstrating strong representational power. Notably, both the \textit{Compressed} and \textit{Quantized Features} yield significant accuracy gains over raw inputs, while substantially reducing dimensionality.
This efficiency is largely attributed to WiFo-CF’s pretraining on the CSI feedback task, which enables the extraction of compact yet informative CSI representations. In contrast to LWM, which adopts masked modeling, WiFo-CF produces representations that are not only more computationally efficient but also better aligned with downstream task requirements—supporting flexible and lightweight deployment across a wide range of edge devices.
}

\subsubsection{Ablation Experiments}

\begin{table*}[h]
\centering
\renewcommand\arraystretch{1.2}  
\caption{Test results of ablation experiments for network architecture and the pre-training strategy.}
\begin{tabular}{c|c|c|c|c|c}
\toprule
\textbf{Metric} & \textbf{WiFo-CF-Base} & \textbf{w/o routed expert} & \textbf{w/o shared expert} & \textbf{w/o multi user} & \textbf{w/o multi rate} \\ \hline
\textbf{NMSE (dB) on Q1-Q8} & \textbf{-4.131}  & -2.896 (↑1.23) & -3.414 (↑0.72) & -2.861 (↑1.27) & -2.844 (↑1.29) \\ \hline
\textbf{NMSE (dB) on Q9-Q16} & \textbf{-3.243} & -2.204 (↑1.04) & -2.541 (↑0.70) & -2.374 (↑0.87) & -1.736 (↑1.51) \\ 
\bottomrule
\end{tabular}
\vspace{-1em}
\label{tab:ablation}
\end{table*}

To validate the effectiveness of the proposed scheme, we conduct ablation studies on both the network architecture and the pre-training strategy. For the architectural components, we remove the HAP module (\textit{w/o HAP}), the shared experts in the S-R MoE module (\textit{w/o shared experts}), and the routed experts (\textit{w/o routed experts}). For the pre-training strategy, we disable dynamic rate adaptation by using a single fixed rate (\textit{w/o multi-rate}), and replace multi-user joint decoding with single-user parallel processing (\textit{w/o multi-user}). As shown in Table~\ref{tab:ablation}, all ablated variants exhibit performance degradation on both in-distribution and out-of-distribution test sets compared to the standard configuration, highlighting the critical role of each component in enabling the effectiveness of WiFo-CF.

\subsubsection{Hyperparameter Analysis on Model Size, Dataset Scale, and Patch Size}

\begin{figure}[htbp]
    \centering
    \vspace{-1em}
    \includegraphics[width=0.8\linewidth]{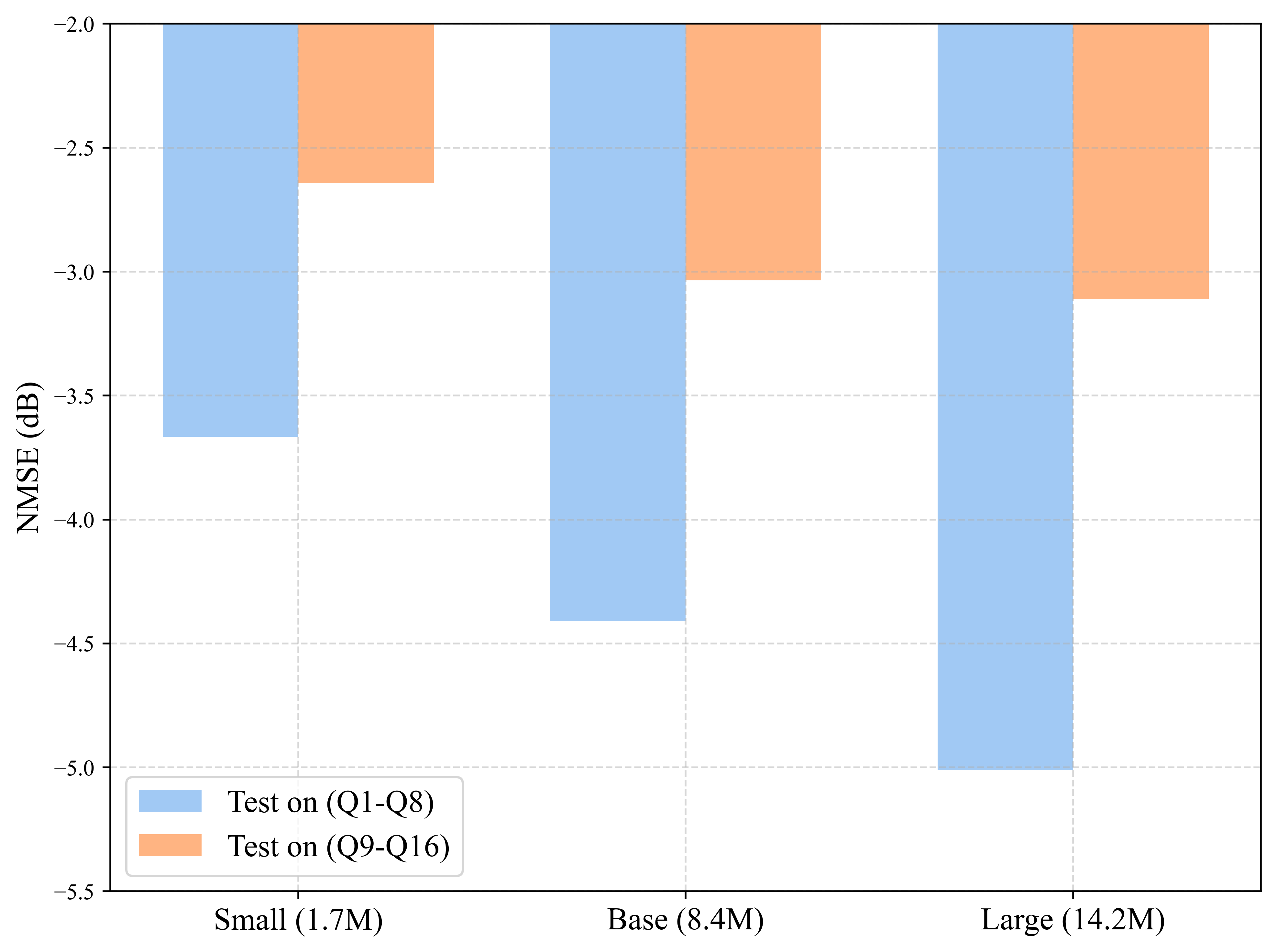}
    \caption{Impact of model size on model performance.}
    \label{fig:Scaling Law model size}
\end{figure}

\begin{figure}[htbp]
    \centering
    \vspace{-1em}
    \includegraphics[width=0.8\linewidth]{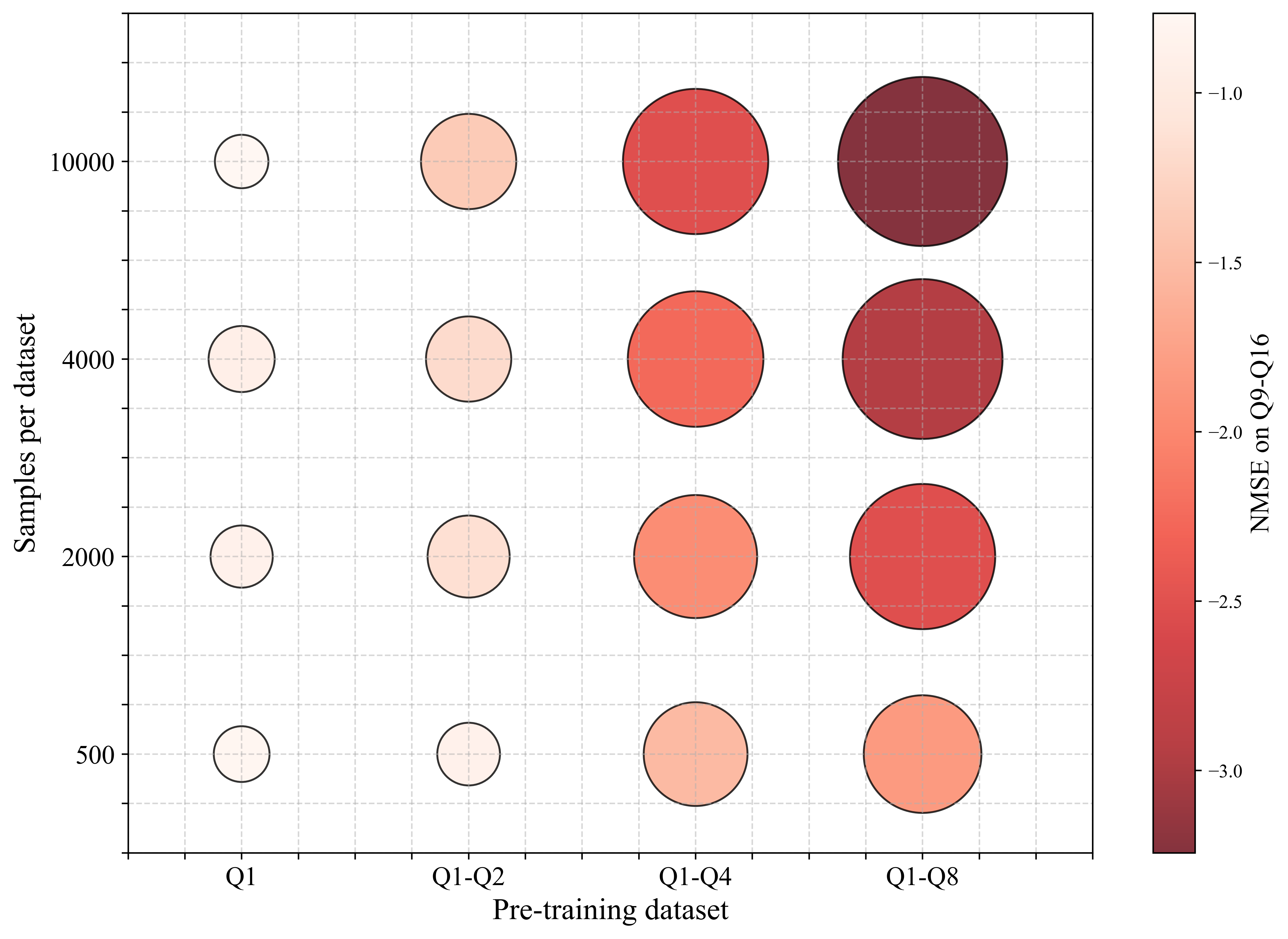}
    \caption{Impact of dataset scale on model performance.}
    \label{fig:Scaling Law dataset}
\end{figure}

We analyze the impact of model size and dataset scale on performance, given the fundamental reliance of wireless foundation models on large-scale data and high-capacity architectures. As shown in Fig.\ref{fig:Scaling Law model size} and Fig.\ref{fig:Scaling Law dataset}, increasing model size consistently improves both in-distribution accuracy and out-of-distribution generalization, due to the enhanced representational capacity. Similarly, expanding the dataset (especially increasing the number of distinct datasets) yields notable gains in generalization, underscoring the critical role of data diversity.

In summary, the success of WiFo-CF stems from both a carefully curated large-scale heterogeneous dataset, which facilitates the learning of generalized channel representations, and a high-capacity model architecture that enables precise modeling of channel characteristics. This advantage is expected to persist and further scale as the dataset and model size continue to grow, reinforcing the practical potential of WiFo-CF.

In addition, the choice of patch size has a noticeable effect on model performance. Larger patches are effective in capturing strong spatial correlations and stationary patterns, while smaller patches offer finer granularity and improved adaptability to local variations. As shown in Table~\ref{tab:patchsize_nmse}, we evaluate the model under different patch size configurations and select $(4, 4)$ in the frequency and antenna domains to strike a balance between local feature extraction and stable pattern modeling.

\begin{table}[h]
\renewcommand\arraystretch{1.3}  
\caption{Model performance under different patch sizes}
\label{tab:patchsize_nmse}
\centering
\scriptsize
\begin{tabular}{c|c|c}
\hline
Patch Size & NMSE on Q1 & FLOPs (M) \\ \hline
(2, 2)  & -6.843 & 525.1  \\ \hline
(2, 4) & -7.257 & 493.7 \\ \hline
(2, 8) & -7.331 & 478.0 \\ \hline
(4, 2) & \underline{-7.550} & 493.7 \\ \hline
\rowcolor{gray!20} 
(4, 4)  & \textbf{-7.624} & 478.0  \\ \hline
(4, 8)  & -6.521 & \underline{470.3} \\ \hline
(8, 2)  & -6.922 & 478.0 \\ \hline
(8, 4)  & -6.075 & \underline{470.3} \\ \hline
(8, 8)  & -4.570 & \textbf{466.6} \\ \hline
\end{tabular}

\end{table}

\subsubsection{Efficiency and Complexity Analysis}

We evaluate the efficiency and complexity of different models from four key perspectives: NMSE performance, the number of inference/training parameters, FLOPs, and inference time, as summarized in Table~\ref{tab:efficiency_comparison}. These metrics are crucial for practical deployment in wireless systems, especially in base station (BS) scenarios with stringent computational and latency constraints.
Despite having only \textbf{80.30M} FLOPs and \textbf{0.10M} inference parameters, WiFo-CF-Small achieves a relatively low NMSE of {-3.667 dB}, demonstrating strong performance-to-complexity trade-off. Compared to its dense counterpart (WiFo-CF-Small (Dense)) with over \textbf{837M} FLOPs and \textbf{1.71M} inference parameters, the sparse version reduces computational cost by over \textbf{90\%} while maintaining comparable accuracy. This highlights the advantage of sparse expert design in reducing redundancy without compromising performance.
Notably, the inference time of WiFo-CF-Small is comparable to that of lightweight baselines such as CRNet and CsiNet. Overall, WiFo-CF-Small achieves an excellent balance among accuracy, parameter efficiency, and inference cost, making it highly suitable for real-time and resource-constrained wireless deployments. Its sparse mixture-of-experts design significantly reduces computational overhead while retaining the performance advantages of a large pretrained model, demonstrating strong potential for practical large-scale applications.

\begin{table*}[t]
\centering
\caption{Efficiency and Complexity Comparison}
\begin{tabular}{lccccc}
\toprule
\textbf{Methods} & \textbf{WiFo-CF-Small} & \textbf{WiFo-CF-Small (Dense)} & \textbf{LAM} & \textbf{CRNet} & \textbf{CsiNet} \\
\midrule
NMSE (dB) & -3.667 & -3.761 & -1.934 & -2.538 & -1.683 \\
Inference / Training Parameters (M) & {0.10} / 1.73 & 1.71 / 1.71 & 0.54 / 0.54 & 0.27 / 0.27 & 0.14 / 0.14 \\
Inference Time (ms) & 7.526 & 7.672 & 3.027 & 2.423 & 1.070 \\
FLOPs (M) & {80.30} & 837.28 & 147.33 & 29.11 & 29.07 \\
\bottomrule
\end{tabular}
\label{tab:efficiency_comparison}
\end{table*}

\section{Conclusions}

This paper introduced WiFo-CF, a wireless foundation model-based architecture tailored for channel feedback. The framework efficiently handled diverse configurations, such as varying input channel dimensions and feedback rates. A multi-user, multi-rate pre-training scheme enhanced the model's ability to learn generalizable channel compression capabilities. The sparse S-R MoE architecture captured shared correlations and unique characteristics across heterogeneous datasets. To support large-scale training and robust generalization, the LH-CDF dataset was developed. Experiments demonstrates that WiFo-CF, pretrained on LH-CDF, excelled in both in-distribution and out-of-distribution tasks and supported downstream channel tasks through dynamic channel representations. Ablation studies confirmed the effectiveness of each component, while complexity analysis highlighted the computational benefits of the sparse design, indicating its potential for scalable and efficient deployment in future wireless systems.

\bibliographystyle{IEEEtran}
\bibliography{WiFo-CF}

\end{document}